\documentclass[aps,prd,groupedaddress,nofootinbib,amssymb,balancelastpage,preprintnumbers]{revtex4}
\usepackage[dvipdfmx]{graphicx}
\usepackage{graphicx,bm,color}
\usepackage{subfigure}
\usepackage{amsmath}
\usepackage{amssymb}
\usepackage{amsfonts}
\usepackage{cases}
\usepackage{cancel}
\usepackage{hyperref}
\usepackage{ulem}

\usepackage{ascmac}

\usepackage{here}
\usepackage{tikz}
\usetikzlibrary{matrix}

\begin{document}

\title{
Effect of quark anomalous magnetic moment on neutral dense quark matter under magnetic field
}

\author{Mamiya Kawaguchi}\thanks{{\tt mamiya@aust.edu.cn}} 
      \affiliation{ 
Center for Fundamental Physics, School of Mechanics and Physics, Anhui University of Science and Technology, Huainan, Anhui 232001, People’s Republic of China
} 
\author{Irfan Siddique}\thanks{{\tt irfansiddique@ucas.ac.cn}} 
      \affiliation{ 
School of Nuclear Science and Technology, University of Chinese Academy of Sciences, Beijing 100049, China
} 

\author{Mei Huang}\thanks{{\tt huangmei@ucas.ac.cn}} 
      \affiliation{ 
School of Nuclear Science and Technology, University of Chinese Academy of Sciences, Beijing 100049, China
}

\begin{abstract}

We discuss the effect of the quark anomalous magnetic moment (AMM) on the neutral dense quark matter under magnetic fields based on the Nambu-Jona-Lasinio (NJL) model at finite baryon density. To address its correlation with the chiral symmetry, we consider a simplified  situation: the model includes the two-quark flavors under constant magnetic fields, and incorporates the effective interaction of the quark AMM linked to the spontaneous chiral symmetry breaking. We then examine the equation of state (EoS) in cases with and without magnetization for anatomizing the thermodynamic quantities. Without the magnetization, a small magnetic field stiffens the EoS, but with increasing the magnetic field, the EoS tends to soften. The stiffness of the EoS is found to be influenced by the magnetic effect on the critical chemical potential of the chiral phase transition and the quark number density at this critical point. As a result, the mass and radius of the neutral dense quark matter increase with the small magnetic field but turn to decrease as the magnetic field further increases. By including the quark AMM, the critical chemical potential is decreased and the quark number density takes a smaller value. Thus, for the stronger magnetic fields, the quark AMM suppresses the softening effect of the magnetic field on the EoS, leading to increased mass and radius compared to when the quark AMM is absent. In contrast, for the small magnetic field, the contribution of the quark AMM to the EoS is marginal. When the magnetization is taken into account, the magnetic effect on the stiffness of the EoS is overshadowed by the contribution of the magnetization, and the significance of the quark AMM also becomes invisible in the mass-radius relation.

\end{abstract}

\maketitle

%%%%%%%%%%%%%%%%%%%%%%%%%%%%%%%%%%%%%%%%%%%%%%%%%%%%%%%%%%%%%%%%%
\section{Introduction}

Investigating the properties of magnetized Quantum Chromo-Dynamics (QCD) in extreme conditions, such as high temperature and high density, is an essential topic for understanding phenomena in non-central relativistic heavy ion collisions and neutron stars, where magnetic fields are present~\cite{Duncan:1992hi,Skokov:2009qp,Voronyuk:2011jd,Deng:2012pc}.
To investigate finite temperature systems relevant to the heavy ion collisions, first principle calculations based on lattice QCD formalism have been used and have provided nonperturbative insights into thermomagnetic QCD, especially the magnetic effect on the QCD phase transitions~\cite{Bali:2011qj,Bali:2012zg,Bali:2013esa,Bornyakov:2013eya,Bruckmann:2013oba,DElia:2018xwo,Endrodi:2019zrl,Tomiya:2019nym,DElia:2021yvk}. To elucidate the lattice QCD observations, the effective model analysis based on the chiral symmetry of the underlying QCD theory has also been employed, such as the Nambu-Jona-Lasinio (NJL) model~\cite{Klevansky:1989vi,Klimenko:1990rh,Gusynin:1995nb} and the chiral perturbation theory~\cite{Andersen:2012zc}.
On the other hand, turning our focus to finite dense systems related to neutron stars, the lattice QCD simulations are difficult to apply at large quark chemical potentials due to the sign problem~\cite{Aarts:2015tyj,Nagata:2021ugx}.
Compared to finite temperature systems,
our understanding of magnetic effects on the high-density matters is limited.

To address the properties of the high densities under the magnetic field, 
insights and findings from the thermomagnetic QCD might be useful. 
At finite temperatures in the absence of the quark chemical potential,  
the magnetic effect on the chiral phase transition has been extensively studied by using the lattice QCD simulations and the effective models. 
Indeed, the lattice QCD simulations have shown that the magnetic field decreases the pseudocritical temperature of the chiral phase transition and reduces the chiral symmetry breaking across the pseudocritical temperature~\cite{Bali:2011qj,Bali:2012zg,Bali:2013esa}. However, the conventional effective model, such as the NJL model, have shown that the chiral symmetry breaking is enhanced by the magnetic field and the pseudocritical temperature increases as well.
This discrepancy between lattice QCD simulations and effective models indicates the need for additional contributions or interactions in the effective model analysis.
In fact, some mechanisms have already been proposed, such as pion fluctuation~\cite{Fukushima:2012kc,Mao:2016fha}, chirality imbalance~\cite{Chao:2013qpa,Yu:2014sla}, and intrinsic magnetic-field dependence on coupling constant~\cite{Ferrer:2014qka,Farias:2016gmy,Coppola:2018vkw,Endrodi:2019whh}.

As another mechanism,
the quark anomalous magnetic moment (AMM) has also become a focus of attention through thermomagnetic studies~\cite{Fayazbakhsh:2014mca,Chaudhuri:2019lbw,Chaudhuri:2020lga,Mei:2020jzn,Xu:2020yag,Ghosh:2020xwp,Ghosh:2021dlo,Farias:2021fci,Chaudhuri:2021skc,Lin:2022ied,Kawaguchi:2022dbq,Mao:2022dqn,Chaudhuri:2022oru}.
In the perturbative framework of massless quarks, the contribution of the quark AMM is forbidden by the chiral symmetry. However, when the chiral symmetry is spontaneously broken in the low energy regime of QCD, quarks would acquire the AMM. As investigated in Ref.~\cite{Chang:2010hb}, it has been shown that spontaneous chiral symmetry breaking dynamically generates the AMM, even for massless quarks.
Additionally, the connection between the quark AMM and the spontaneous chiral symmetry
breaking has also been studied within the NJL model framework~\cite{Ghosh:2021dlo}.
These studies indicate that the quark AMM would depend on the order parameter of the spontaneous chiral symmetry breaking which is known as the chiral condensate.
Then, it has been anticipated that the quark AMM serves as an additional contribution to bridge the gap between lattice QCD simulations and effective models.
However, the explicit expression of the quark AMM
is still unknown.

To explore its expression,
the role of the quark AMM has been investigated in the NJL model including its effective interaction in the thermomagnetic system~\cite{Kawaguchi:2022dbq}. It has been found that the effective interaction of the quark AMM significantly influences the order of the chiral phase transition. 
By utilizing this characteristic, the form of the quark AMM has been restricted to satisfy the chiral crossover observed in
the lattice QCD simulations~\cite{Kawaguchi:2022dbq}. Eventually, an appropriate form has been proposed, which is proportional to the square of the chiral condensates.
More precisely, this proposed effective interaction of the quark AMM is described solely by chiral condensates, magnetic fields and quark fields. Hence, its effective interaction is adaptable to the finite quark chemical potential system, which would be a clue for exploring the properties of the high-density matters.

Namely, given the link between the quark AMM and the spontaneous chiral symmetry breaking, the proposed effective form of the quark AMM is expected to be significant in high-density matters associated with neutron stars across the chiral phase transition. In this study, we investigate the role of the quark AMM in the neutral dense quark matter under the magnetic fields.
Although several studies have used the effective models for investigating the properties of the neutral dense quark matter under the magnetic field~\cite{Menezes:2009uc,Chu:2014foa,Chu:2014pja}, but the quark AMM has not been taken into account.
Note that while the effective form of the quark AMM has been applied to systems with finite quark chemical potential at zero temperature based on the NJL model~\cite{He:2022inw}, this previous study did not address neutral dense quark matter and thermodynamic properties such as pressure, energy and magnetization have not been considered.
In our study, to clearly grasp the role of the quark AMM in the neutral dense quark matter, we consider an ideal situation based on a simple model: the two-flavor NJL model including the quark AMM under external constant magnetic fields. 
Within the mean field approximation in the NJL model, while satisfying the $\beta$ equilibrium and the electric charge neutrality, 
we then evaluate the thermodynamic properties.
Furthermore, the mass and radius relation of the neutral dense quark matter is also addressed to better understand the significance of the quark AMM in the high-density matters under the magnetic field.

\section{Model including the quark AMM}

In this study, we aim to delve into the understanding
of the effect of the quark AMM in the high-density matter at zero temperature.
To examine the neutral dense quark matter under the magnetic field, we employ a simple effective model described by the following Lagrangian,
\begin{eqnarray}
{\cal L} = 
{\cal L}_{\rm q} + {\cal L}_{\rm l},
\end{eqnarray}
where ${\cal L}_{\rm q}$ and ${\cal L}_{\rm l}$ represent parts for the quarks and leptons, respectively.
The Lagrangian for quarks is based on the two-flavor NJL model, which includes the effective interaction of the quark AMMs, 
\begin{eqnarray}
{\cal L}_{\rm q}
&=&
\sum_{f=u,d} \bar q^f 
(i\gamma^\mu  D_\mu^f  -m_f )q^f 
+{\cal L}_{\rm int }^{(4)}
+{\cal L}^{\rm (AMM)}_{\rm int},
\label{NJL}
\end{eqnarray}
where $q^f$ is the two-flavor quark field, $q=(u,d)^T$; the subscript $f$ denotes the quark flavor; 
$m_f$ are the current quark mass.
The electromagnetic gauge field is embedded into 
the covariant derivative, $D_\mu^f=\partial_\mu -iQ _fA_\mu -i\mu_f\, \delta_{\mu0}$, with the electric charge for quark flavor $Q_u = +2e/3$ and $Q_d = -e/3$.
In this study, we consider constant magnetic fields along the $z$-direction in four-dimensional spacetime,
$A_\mu = (0, -By/2, Bx/2,0)$.
In addition,
the chemical potential for each quark flavor $\mu_f$ are also embedded in the covariant derivative, which can be rewritten in terms of the quark chemical potential $\mu_q$ and the isospin chemical potential $\mu_I$,
\begin{eqnarray}
\mu_u &=& \mu_q +\frac{1}{2}\mu_I,\nonumber\\
\mu_d &=& \mu_q -\frac{1}{2}\mu_I.
\end{eqnarray}
The second term in Eq.~(\ref{NJL}) describes as the four-point interactions of quarks in the scalar and pseudoscalar channels,
\begin{eqnarray}
{\cal L}_{\rm int }^{(4)} = G_S\left\{
(\bar qq)^2 +(\bar q i\gamma_5\vec \tau q)^2
\right\},
\end{eqnarray}
with the coupling constant $G_S$ and $\vec \tau$ being the Pauli matrix. 
The third term in Eq.~(\ref{NJL}) expresses the effective interaction of the quark AMM, 
\begin{eqnarray}
{\cal L}^{\rm (AMM)}_{\rm int}
&=&
\sum_{f=u,d}
\frac{1}{2} \kappa_f Q_f
F_{\mu\nu}
\bar q^f \sigma^{\mu\nu} q^f,
\label{int_AMM}
\end{eqnarray}
where 
$F_{\mu\nu}= \partial_\mu A_\nu -\partial_\nu A_\mu$ is the field strength of $A_\mu$; $\sigma_{\mu\nu} = \frac{i}{2}[\gamma_\mu,\gamma_\nu]$;
$\kappa_f$ is the dimensionful coupling and represents the quark AMM. 
The detailed expression of $\kappa_f$ will be discussed later.

As for the lepton sector, 
the Lagrangian ${\cal L}_{\rm l}$ is described by
\begin{eqnarray}
{\cal L}_{\rm l} = \bar e
\left( i\gamma^\mu\partial_\mu
+e\gamma^\mu A_\mu
+\mu_e \gamma^0
\right)
e
+
\bar \mu
\left(
i\gamma^\mu  \partial_\mu
+e\gamma^\mu A_\mu
+\mu_\mu \gamma^0
-m_\mu
\right)
\mu,
\end{eqnarray}
where $e$ and $\mu$ denote the electron and muon fields respectively; $\mu_e$ and $\mu_{\mu}$ represent their respective chemical potentials;
$m_\mu= 105.66 {\rm MeV}$ is the muon mass.

%%%%%%%%%%%%%%%%%%%%%%%%%%%%%%%%%%%%%%%%%%%%%%%%%%%
\subsection{Effective potential including the quark AMM}

To describe the neutral dense quark matter across the chiral phase transition,
in this study, we adapt the mean-field approximation for the four point interaction term of quarks,
\begin{eqnarray}
{\cal L}_{\rm int }^{(4)} = G_S\left\{
(\bar qq)^2 +(\bar q i\gamma_5\vec \tau q)^2
\right\}
&\to&
4G_S\langle \bar u u \rangle \bar u u
+
4G_S\langle \bar dd \rangle \bar dd
-2G_S  \langle \bar u u \rangle^2
-2G_S  \langle \bar dd \rangle^2.
\end{eqnarray}
In connection with this, the scalar mean fields are introduced as follows:
\begin{eqnarray}
\sigma_f = - 4G_S  
\langle \bar q_f q_f
\rangle.
\end{eqnarray}
Within the mean-field approximation,
the NJL Lagrangian is of the form,
\begin{eqnarray}
{\cal L}_{\rm q}^{\rm mean} &=&
\sum_{f=u,d}
\left[
\bar q^f \left(%i\gamma^\mu D_\mu 
i\gamma^\mu \partial_\mu +Q_f \gamma^\mu A_\mu +\mu_f 
\gamma^0
-M_f+\frac{1}{2}\kappa_{f} Q_f F_{\mu\nu} \sigma^{\mu\nu} \right)q^f
-\frac{1}{8G_S}\sigma_f^2
\right],
\end{eqnarray}
where $M_f =m_f +\sigma_f$ are
the dynamical quark masses. By integrating out the quark and lepton fields in the generating functional,
we obtain the effective potential at zero temperature, which includes the quark AMMs,
\begin{eqnarray}
V_{\rm eff} (\mu_u, \mu_d, \mu_e, \mu_\mu,eB; \sigma_u,\sigma_d)
%&=&\sum_{f=u,d}\left[\frac{1}{8G_S}\sigma^2_f -N_c|Q_f B|\sum_{l=0}^\infty\sum_{s=\pm1}\int_{-\infty}^\infty\frac{dp_3}{4\pi^2}\left\{E_{f}^{(l,s)}-\left(E_{f}^{(l,s)}-\mu_f\right) \theta \left(\mu_f -  E_{f}^{(l,s)}\right)\right\}\right]\nonumber\\
&=&
V^{\rm S}_{\rm eff} + V^{\rm loop}_{\rm eff},
\label{V_eff}
\end{eqnarray}
where 
$V^{\rm S}_{\rm eff}$ is
the scalar potential part described by the mean fields and $V^{\rm loop}_{\rm eff}$ is driven by the loop calculations of quarks and leptons,
\begin{eqnarray}
V^{\rm S}_{\rm eff}
&=&
\sum_{f=u,d}
\frac{1}{8G_S}\sigma^2_f
,\nonumber\\
V^{\rm loop}_{\rm eff}&=&
-\sum_{f=u,d}
N_c
|Q_f B|
\sum_{l=0}^\infty
\sum_{s=\pm1}
\int_{-\infty}^\infty
\frac{dp_3}{4\pi^2}
\left\{
E_{f}^{(l,s)}
f_{\Lambda,f}^{(l,s)}
-
\left(E_{f}^{(l,s)}-\mu_f\right) \theta \left(\mu_f -  E_{f}^{(l,s)}\right)
\right\}\nonumber\\
&&
-
|e B|
\sum_{l=0}^\infty
\sum_{s=\pm1}
\int_{-\infty}^\infty
\frac{dp_3}{4\pi^2}
\left[
\left\{
-
\left(E_{e}^{(l,s)}-\mu_e\right) \theta \left(\mu_e-  E_{e}^{(l,s)}\right)
\right\}
+
\left\{
-
\left(E_{\mu}^{(l,s)}-\mu_\mu\right) \theta \left(\mu_\mu  -  E_{\mu}^{(l,s)}\right)
\right\}
\right],
%E_{f}^{(l,s)}=
%\sqrt{
%p_3^2+
%\left[
%\left\{
%|Q_f B|(
%2l+1 -s \xi_f
%)+M_f^2
%\right\}^{1/2}
%-s \kappa_fQ_f B
%\right]^2
%},
\label{potential_detail}
\end{eqnarray}
with 
$N_c$ denoting 
the number of colors,
$l$ representing the landau level, 
$s=\pm 1$ representing the spin up/down of the quarks and $E_{i}^{(l,s)}$ being
the energy dispersion relations
\begin{eqnarray}
E_{i}^{(l,s)}&=&
\sqrt{
%\left(
p_3^2+
\left({M}_{{\rm eff},i}^{(l,s)}\right)^2
},
\;\;\;
({\rm for}\;\;\;i=u,d),
\nonumber\\
E_{i}^{(l,s)}&=&
\sqrt{
%\left(
p_3^2+
\left({m}_{{\rm eff},i}^{(l,s)}\right)^2
},
\;\;\;
({\rm for}\;\;\;i=e,\mu).
\end{eqnarray}
Here, $p_3$ denotes the third component in the momentum
space and
$M_{{\rm eff},i}^{(l,s)}$ represents the effective mass of quarks and
$m_{{\rm eff},i}^{(l,s)}$ denotes the effective mass of leptons under the magnetic field,
\begin{eqnarray}
M_{{\rm eff},f}^{(l,s)}&=&
\left\{
|Q_f B|(
2l+1 -s \xi_f
)+M_f^2
\right\}^{1/2}
-s \kappa_fQ_f B,\nonumber\\
m_{{\rm eff}, e}^{(l,s)} &=&
\sqrt{|e B|(2l+1 +s )},\nonumber\\
m_{{\rm eff}, \mu}^{(l,s)}
&=&
\sqrt{|e B|(
2l+1 +s 
)+m_\mu^2},
\end{eqnarray}
with $\xi_f={\rm sgn}(q_fB)$.

The loop part of the effective potential in Eq.~(\ref{V_eff}) involves
an ultraviolet (UV) divergence.
To regularize this divergence, 
we insert %employ a smooth regularization scheme by incorporating 
the regulator function $f_{\Lambda,f}^{(l,s)}$ into the effective potential as described in Eq.~(\ref{potential_detail}).
In this study, we adopt the Lorentzian form factor for the regulator function $f_{\Lambda,f}^{(l,s)}$ as follows:
\begin{eqnarray} 
f_{\Lambda,f}^{(l,s)}
=
\frac{\Lambda^{10}}{\Lambda^{10}+(\sqrt{p_3^2 +|q_f B|(2l+1-s\xi_f )})^{10}
},
\label{regulator}
\end{eqnarray}
where $\Lambda$ denotes the ultraviolet momentum cutoff.
Note that the loop part of the leptons in the effective potential also has the UV divergence, but this divergence part is ignored in Eq.~(\ref{potential_detail}) because it becomes irrelevant when considering the physical thermodynamic quantities such as subtracted pressure and subtracted energy which are used to determine the EoS (more details are discussed in the following). 
%\textcolor{red}{
As for the choice of the regularization scheme, the magnetic field independent regularization (MFIR) is the preferred approach to avoid the nonphysical oscillation behavior in the quark condensate under magnetic fields~\cite{Avancini:2019wed}. 
%However, once we include the quark AMM, the Landau-level representation in the quark propagator becomes intricate, making it difficult to apply the conventional MFIR method.  Therefore, we have chosen the Lorentzian form-factor regularization.
Although it has been reported in \cite{Avancini:2019wed} that the Lorentzian form-factor regularization leads to oscillations in the quark condensate,
the significant oscillations do not appear in the subtracted quark condensate at zero and finite temperatures as shown in the previous study~\cite{Kawaguchi:2022dbq}.
While a slightly wavy shape appear in 
the previous result, removing this minor wavy behavior would not crucially affect our current results based on the mean field analysis.
This is because this minor behavior
is within the permissible range of the large $N_c$ expansion in the NJL model analysis.
%}

With the UV regularization, one can determine the expectation value of the scalar mean field from the stationary condition of the effective potential,
\begin{eqnarray}
\frac{\partial V_{\rm eff}}{\partial \sigma_f  } &=&0.
\label{stationary}
\end{eqnarray}
The determined expectation values correspond to the order parameter of the spontaneous chiral symmetry breaking, $\langle \bar q q \rangle\sim \sum_f\sigma_f$, which is called the chiral condensate.
\\

Now, we provide a detail of the quark AMM. 
As a reference for considering the quark AMM, we first  take a look at the case of QED, which is described by the electron and dynamical photon fields. Within the perturbative framework of QED, 
the electron AMM is driven by loop contributions for the photon-electron-antielectron vertex function.
Referring to the one-loop evaluation of the electron AMM, 
one would roughly estimate
the quark AMM as $\kappa_f = {Q_f^2}/({16\pi^2 m_f})$ at the one-loop level. 
However, the Bethe-Salpeter approach following QCD indicates that
the spontaneous chiral symmetry breaking generates
the quark AMM even for massless current quarks~\cite{Chang:2010hb}.
Namely, 
in the low-energy regime of QCD,
the quark AMM is closely connected with the nonperturbative phenomenon of the chiral dynamics rather than the perturbative effect of the one-loop evaluation.
In addition, the connection between the quark AMM and the spontaneous chiral symmetry breaking has also been mentioned in the two-flavor NJL model within the mean field approximation~\cite{Ghosh:2021dlo}, where the four-point interaction is incorporated in the loop calculation for the photon-quark-antiquark vertex function.
These studies suggest that the quark AMM depends on the chiral condensate: $\kappa_f(\sigma_f)$.
However, the specific form of the quark AMM associated with the spontaneous chiral symmetry breaking is still unknown.

To clarify the role of the quark AMM in the chiral dynamics, several studies have investigated its effect by using the NJL model analysis at finite temperature, and found that 
the effective interaction of
the quark AMM affects the order of the chiral phase transition in the thermomagnetic system~\cite{Fayazbakhsh:2014mca,Chaudhuri:2019lbw,Chaudhuri:2020lga,Mei:2020jzn,Xu:2020yag,Ghosh:2020xwp,Ghosh:2021dlo,Farias:2021fci,Chaudhuri:2021skc,Lin:2022ied,Kawaguchi:2022dbq,Mao:2022dqn,Chaudhuri:2022oru}.
For simplicity, when a constant AMM is assumed, it leads to an unexpected first order phase transition, which is inconsistent with the chiral crossover observed in the lattice QCD simulations. 
%\textcolor{red}{
In a recent study~\cite{Tavares:2023oln}, it has been reported that 
the unexpected first-order phase transition, observed in the zero-temperature system, could be caused by regularization schemes.
On the other hand, an effective form of the quark AMM has also been proposed to achieve better qualitative agreement with lattice QCD simulations at finite temperature~\cite{Lin:2022ied,Kawaguchi:2022dbq}:
%}
%To achieve better qualitative agreement with lattice QCD simulations, an effective form of the quark AMM has been  proposed:
\begin{eqnarray}
\kappa_f (\sigma_f) &=&\bar v_f \sigma_f^2,
\label{AMM_form}
\end{eqnarray}
where $\bar v_f$ is a dimensionful model parameter 
called the AMM parameter. 
Even with the use of this AMM, the chiral crossover is maintained in the thermomagnetic system, and it has the effect of suppressing the magnetic enhancement in the chiral condensate.
%\textcolor{red}{
Note that the effective form of the quark AMM in Eq.~(\ref{AMM_form}) is also motivated by the previous study of electron AMM in the QED framework under the magnetic field. Reference~\cite{Lin:2021bqv} investigated the external magnetic field contribution to the AMM for the electron, which is proportional to the electron mass squared
with magnetic corrections. From this fact, we naively extended the QED framework to the QCD case, promoting the electron mass to the dynamical quark mass.
With $M_f\sim \sigma_f$, we have arrived the assumption of the quark AMM, $\kappa_f  \sim \sigma_f^2$.
%}

Recently, the effective form in Eq.~(\ref{AMM_form}) has been applied to the systems with finite quark chemical potential at zero temperature~\cite{He:2022inw}. 
In this previous study, the properties of the quark matter under the magnetic field is investigated, but the neutral dense quark matter were not addressed and the flavor symmetry breaking was not fully accounted.
In this study, we apply the effective form in Eq.~(\ref{AMM_form}) to 
the neutral dense quark matter by using the NJL model analysis with taking the flavor symmetry breaking into account.

%%%%%%%%%%%%%%%%%%%%%%%%%%%%%%%%%%%%%%%%%%%%%%%%%%%
\subsection{Thermodynamic quantities}

Using the effective potential in Eq.~(\ref{V_eff}) with the chiral condensates determined from the stationary condition , 
we define the isotropic pressure $p_0$ of the quark matter, which also includes the lepton contribution,
\begin{eqnarray}
p_0&=& - V_{\rm eff}.
\end{eqnarray}
By using this isotropic pressure, the number density of quarks and leptons are evaluated as
\begin{eqnarray}
\rho_f   
%&=& N_c|Q_f B|\sum_{l=0}^\infty\sum_{s=\pm1}\int_{-\infty}^\infty \frac{dp_3}{4\pi^2}\theta \left(\mu_f -  E_{f}^{(l,s)}\right), \nonumber\\
&=&
\frac{\partial p_0  }{\partial \mu_f}
=
N_c
|Q_f B|
\sum_{l=0}^\infty
\sum_{s=\pm1}
\frac{1}{2\pi^2}
\sqrt{\mu_f^2-\left(M_{{\rm eff},f}^{(l,s)}\right)^2 },
\nonumber\\
\rho_e 
%&=& |e B|\sum_{l=0}^\infty\sum_{s=\pm1}\int_{-\infty}^\infty\frac{dp_3}{4\pi^2}\theta \left(\mu_e-  E_{e}^{(l,s)}\right),\nonumber\\
&=&
\frac{\partial p_0  }{\partial \mu_e}
=
|e B|
\sum_{l=0}^\infty
\sum_{s=\pm1}
\frac{1}{2\pi^2}
\sqrt{\mu_e^2-\left(M_{{\rm eff},e}^{(l,s)}\right)^2 },
\nonumber\\
\rho_\mu 
%&=& |e B|\sum_{l=0}^\infty\sum_{s=\pm1}\int_{-\infty}^\infty\frac{dp_3}{4\pi^2}\theta \left(\mu_\mu-  E_{\mu}^{(l,s)}\right)\nonumber\\
&=&
\frac{\partial p_0  }{\partial \mu_\mu}
=
|e B|
\sum_{l=0}^\infty
\sum_{s=\pm1}
\frac{1}{2\pi^2}
\sqrt{\mu_\mu^2-\left(M_{{\rm eff},\mu}^{(l,s)}\right)^2 }.
\end{eqnarray}
Note that the number densities are explicitly free from the UV divergence.
In addition, based on these number density of quarks, the quark number density $\rho_q$ is defined as
\begin{eqnarray}
\rho_q = \frac{\rho_u + \rho_d}{2}.
\end{eqnarray}

Using the pressure and the number densities, 
the energy ($\epsilon_0$) of the dense matter composed of quarks and leptons is also evaluated as
\begin{eqnarray}
\epsilon_0&=& - p_0 + \sum_{i=u,d,e,\mu} \mu_i \rho_i.
\end{eqnarray}
In the vacuum corresponding to the zero chemical potential, the pressure and energy have finite values associated with UV divergence, which should be subtracted to obtain zero value. This subtraction is defined as
\begin{eqnarray}
p_0^{\rm sub} &=& p_0 (\mu_u,\mu_d,\mu_e, \mu_\mu) 
-
p_0 (\mu_u=\mu_d=\mu_e= \mu_\mu=0),\nonumber\\
\epsilon_0^{\rm sub}&=& - p_0^{\rm sub} + \sum_{i=u,d,e,\mu} \mu_i \rho_i.
\end{eqnarray}
These subtracted pressure and energy represent the physical quantities.
\\

In our study, we consider the neutral dense quark matter composed of quarks and leptons, which is described by satisfying the $\beta$ equilibrium and the electric charge neutrality.
In the two-flavor system,
the $\beta$ equilibrium is expressed by
the following chemical equilibrium condition,
\begin{eqnarray}
\mu_u +\mu_e =\mu_d,\;\;\;
\mu_\mu = \mu_e.
\label{beta_equ}
\end{eqnarray}
Furthermore, the electric charge neutrality in the quark matter is maintained by  
\begin{eqnarray}
\frac{2}{3}\rho_u -\frac{1}{3}\rho_d -\rho_e -\rho_\mu =0.
\label{neutrality}
\end{eqnarray}
To address the neutral dense quark matter, we impose two conditions given by Eq.~(\ref{beta_equ}) and Eq.~(\ref{neutrality})
when solving the stationary condition in Eq.~(\ref{stationary}). 
Using the determined chiral condensates,
we then evaluate the thermodynamic quantities of the neutral dense quark matter, such as pressure, energy and magnetization.

%%%%%%%%%%%%%%%%%%%%%%%%%%%%%%%%%%%%%%%%%%%%%%%%%%%
\subsection{Magnetization and anisotropic pressure}

Under the magnetic field, the quark matter exhibits the magnetization ${\cal M}$: 
\begin{eqnarray}
{\cal M} = \frac{\partial p_0 }{\partial B}.
\end{eqnarray}
In fact, when considering the energy-momentum tensor from the generating functional, the magnetization appear in the diagonal component of the energy-momentum tensor~\cite{Ferrer:2010wz}. 
The magnetization then splits the pressure into the longitudinal pressure 
$p_{\parallel}$, which aligns with the magnetic field, and the transverse pressure $p_\perp$, which is perpendicular to the direction of the magnetic field.
These anisotropic pressures caused by the presence of the magnetization
are defined as~\cite{Ferrer:2010wz}
\begin{eqnarray}
p_{\parallel} &=& p_0
-\frac{1}{2}B^2
,\nonumber\\
p_{\perp} &=&
p_0
+\frac{1}{2}B^2
-{\cal M}B.
\label{p_w_magnetization}
\end{eqnarray}
The subtracted anisotropic pressure is defined as
\begin{eqnarray}
p_{\parallel}^{\rm sub} &=&
p_{\parallel}(\mu_u,\mu_d,\mu_e,\mu_\mu) - 
p_{\parallel}(\mu_u=\mu_d=\mu_e= \mu_\mu=0),\nonumber\\
p_{\perp}^{\rm sub} &=&
p_{\perp}(\mu_u,\mu_d,\mu_e,\mu_\mu) - 
p_{\perp}(\mu_u=\mu_d=\mu_e= \mu_\mu=0).
\label{sub_def}
\end{eqnarray}
By using the above definition of subtracted pressures,
the $B^2$ terms in Eq.~(\ref{p_w_magnetization}) vanish since $p_{\parallel}(\mu_u,\mu_d,\mu_e,\mu_\mu)$ and 
$p_{\perp}(\mu_u=\mu_d=\mu_e= \mu_\mu=0)$ are defined at the same value of $eB$.

Note that there is another way to subtract the pressure, as shown in Ref.~\cite{Chu:2014pja}, where the pressure is subtracted by 
$p_{\parallel,\perp}(\mu_u=\mu_d=\mu_e= \mu_\mu=0,\,eB=0)$. However, under this definition, as the magnetic field increases, $p_{\parallel}^{\rm sub}$ becomes negative across all relevant density regions. This negative longitudinal pressure is induced by the $B^2$ term with increasing magnetic field strength.
Consequently, the negative pressure would lead to the collapse of a neutral dense quark matter owing to its own gravity (for a discussion on collapse induced by the magnetic field, also see Refs.~\cite{Chaichian:1999gd,Huang:2009ue}). To investigate the role of the quark AMM while avoiding such a collapse, we adopt the definition in Eq.~(\ref{sub_def}) instead.
In the next section,
using the subtracted longitudinal and perpendicular pressures in Eq.~(\ref{sub_def}), we will define the averaged pressure to solve the well known Tolman-Oppenheimer-Volkoff equation.
We will then show
the mass and radius of the neutral quark matter under the external magnetic field as
the obtained stable solutions.
Further calculation details are provided in the next section.

Furthermore, the magnetization also has the UV divergence. To evaluate the physical quantity, we consider the subtracted magnetization as follows:
\begin{eqnarray}
{\cal M}_{\rm sub} = {\cal M}(\mu_u,\mu_d,\mu_e, \mu_\mu) - {\cal M}(\mu_u=\mu_d=\mu_e= \mu_\mu=0).
\label{sub_magne}
\end{eqnarray}

%%%%%%%%%%%%%%%%%%%%%%%%%%%%%%%%%%%%%%%%%%%%%%%%%%%
\section{Numerical results}

In this section, we numerically investigate the neutral dense quark matter under the external magnetic field based on the NJL model analysis within the mean-field approximation.
For our numerical calculation, we use the following parameter set,
$\Lambda = 681.38{\rm MeV}$,
$m_u=m_d= $4.552{\rm MeV},
$G_S \Lambda^2= 1.86$~\footnote{
In our study we show numerical results for smooth regulation scheme in Eq.~(\ref{regulator}), but we have also checked the for Pauli-Villars regularization scheme. The numerical results from these two regularization scheme are qualitatively consistent. }.
Before presenting our numerical results, here, we discuss the constraint on the quark AMM parameter $\bar{v}_{u,d}$.

In cases where $\bar{v}_u < 0$, we observe negative dynamical quark masses and/or negative number densities regardless of whether $\bar{v}_d$ is positive or negative. Conversely, in cases where $\bar{v}_u > 0$, such unphysical results are not observed. Therefore, in our numerical analysis, we exclude cases that lead to unphysical observable. 

As for the value of $\bar{v}_{u,d}$,
the maximum value of the AMM parameter
is evaluated through the discussion of the potential stability, which is estimated as $\bar{v}_{u,d} \simeq 2 \, {\rm GeV}^{-3}$~\cite{Kawaguchi:2022dbq}. In this study, we consider $\bar{v}_{u,d} = 2 \, {\rm GeV}^{-3}$ as a large value, and $\bar{v}_{u,d} = 1 \, {\rm GeV}^{-3}$ as a smaller value. 
In addition, since understanding of the quark AMM remains limited, there is a possibility that $\bar{v}_d$ could be negative. Thus, we also explore the case of $\bar{v}_u = 2 \, {\rm GeV}^{-3}$ and $\bar{v}_d = -2 \, {\rm GeV}^{-3}$.
With these cases, we will perform numerical calculations for $eB=1$-$4\,{\rm GeV}^2$
%\textcolor{red}{corresponding to 
$B = 1.69$-$6.76\times 10^{19}\,{\rm G}$.
%}
%\textcolor{red}{
Actually, the magnetic fields in our study are far from the expected values in realistic magnetars: the surface magnetic field of magnetars is estimated to be $B \sim 10^{10}{\rm -} 10^{15} \,{\rm G}$~\cite{Haensel:2007yy} while 
the interiors of magnetars may reach $B \sim 10^{18} \,{\rm G}$~\cite{1991ApJ...383..745L}.
Our main aim is to study how much the quark AMM contributes to quark stars under magnetic fields. To
achieve this, we allowed for some deviation from the realistic magnetic field conditions of magnetars.
Therefore, the magnetic field used in our study is somewhat stronger than what might be expected in a real scenario. As a stance of our paper, regardless of whether the contributions of the quark AMM are relevant in reality, we  investigate the quark matter properties in the magnetic field regions
where the contribution of the quark AMM becomes significant.
%}
\\

To clarify the effect of the quark AMM on the high-density matters under the magnetic field, we first evaluate the dependence of the dynamical quark masses on the quark chemical potential in the neutral dense quark matter and identify the critical chemical potential for the chiral phase transition. 
We then assess the thermodynamic quantities such as quark number density, pressure, energy and magnetization.
In this study, to anatomize the EoS, we consider the pressure and energy in cases with and without the magnetization. 
With the evaluated EoSs, we also address the mass and radius relation of the neutral dense quark matter.

%Finally, using the EoSs, we assess the mass and radius relation of the neutron star.
%Next, we investigate the equation of state through pressure and energy.

%%%%%%%%%%%%%%%%%%%%%%%%%%%%%%%%%%%%%%%%%%%%%%%%%%%
\subsection{
The quark AMM contribution to the chiral symmetry breaking in neutral dense quark matter}

\begin{figure}[H] %htbp
\begin{tabular}{cc}
\begin{minipage}{0.5\hsize}
\begin{center}
    \includegraphics[width=8.8cm]{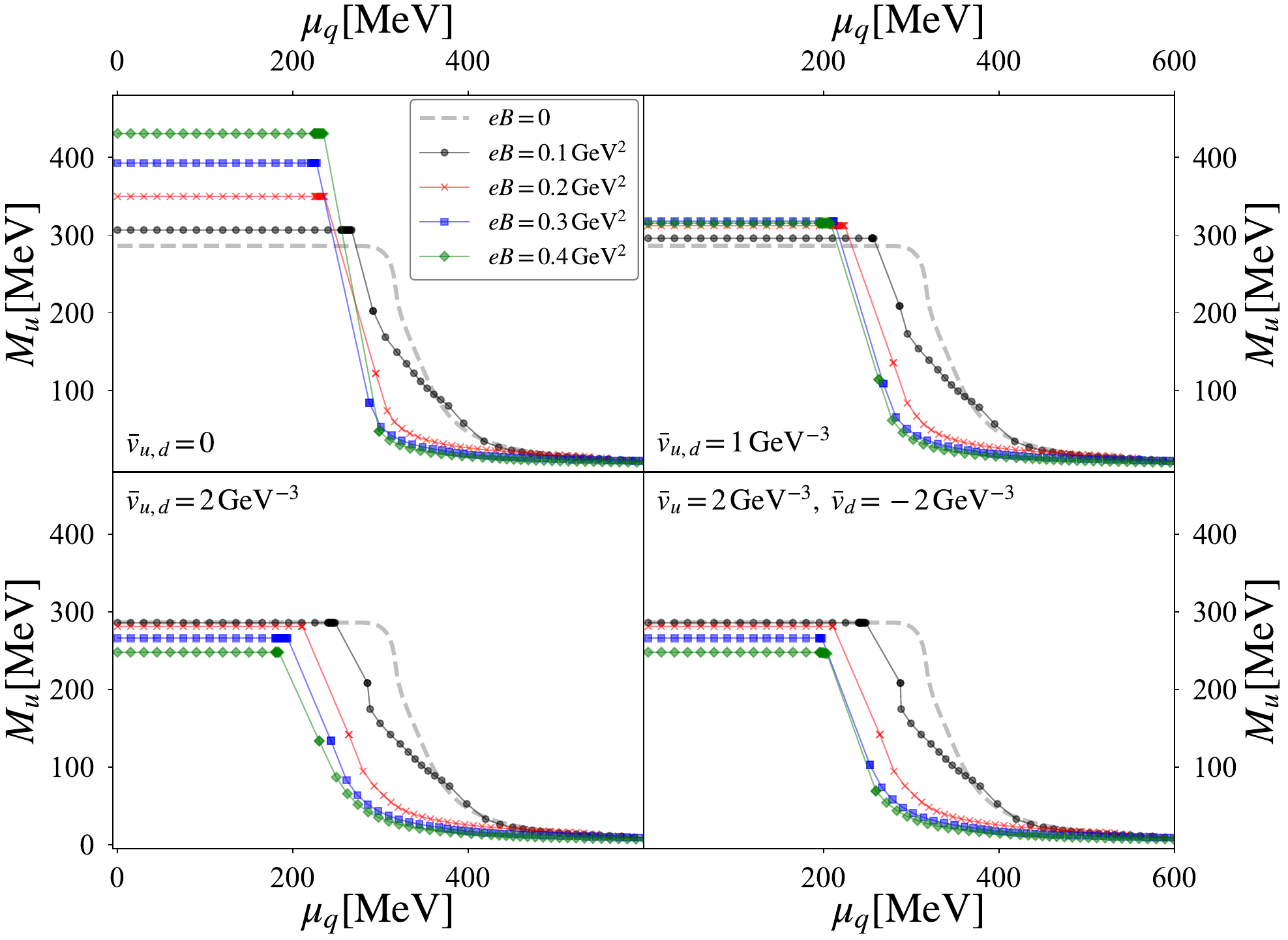}
    \subfigure{(a)}
\end{center}
\end{minipage}
\begin{minipage}{0.5\hsize}
\begin{center}
    \includegraphics[width=8.8cm]{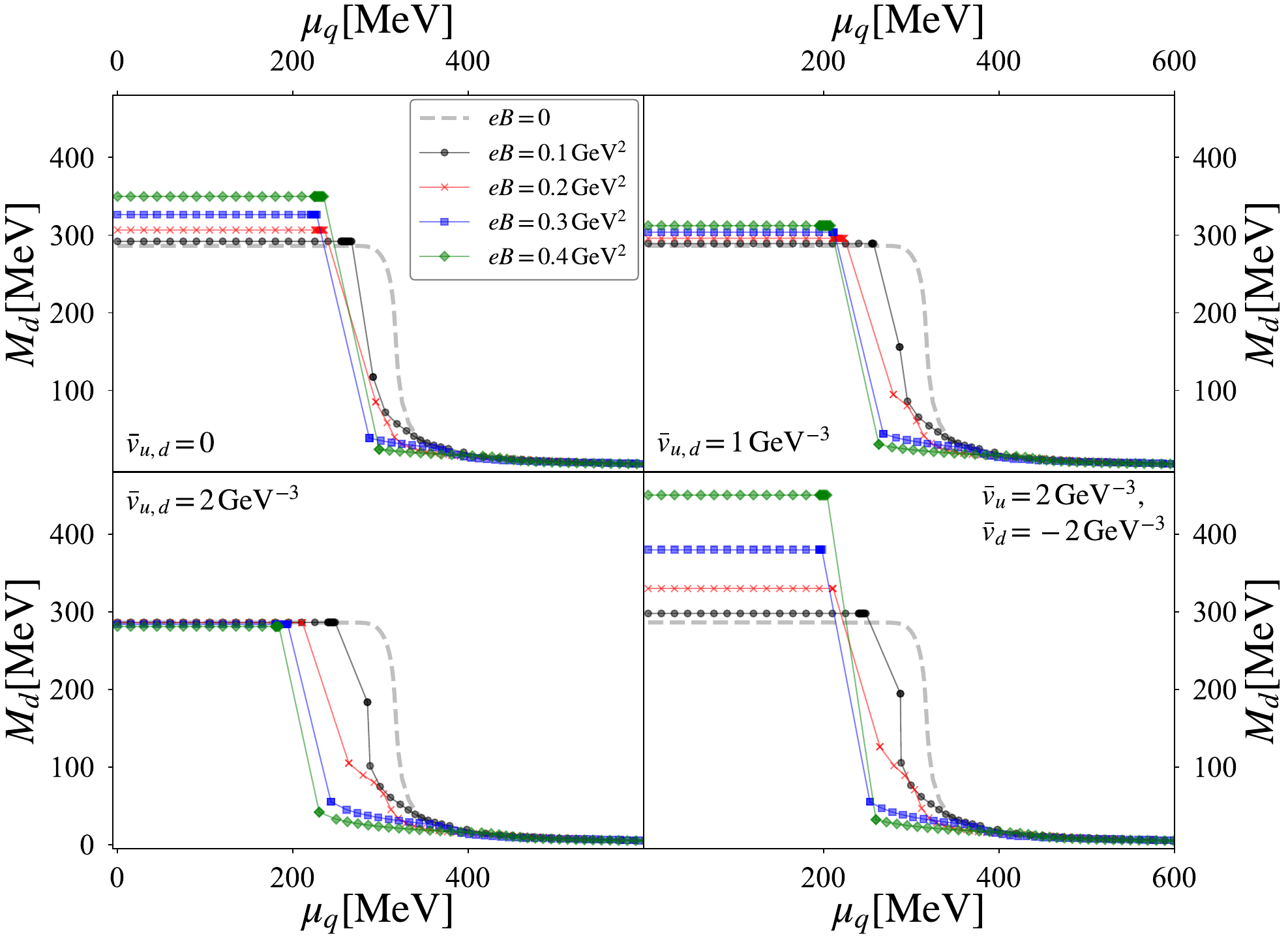}
    \subfigure{(b)}
\end{center}
\end{minipage}
\end{tabular}
\caption{
Quark AMM contribution to dynamical quark masses as the function of the quark chemical potential:
Panel~(a) illustrates the dynamical up-quark mass, while Panel~(b) depicts the dynamical down-quark mass. 
}
\label{mass_chemi}
\end{figure}

In this subsection, we discuss the contribution of the quark AMM to the spontaneous chiral symmetry breaking by monitoring the quark chemical potential dependence of the dynamical quark masses.
We plot the dynamical masses of up (panel~(a)) and down (panel~(b)) quarks as a function of the quark chemical potential in Fig.~\ref{mass_chemi}. This figure shows that there is a jump in the dynamical quark masses for all cases of the quark AMM under the magnetic fields, indicating the first-order chiral phase transition.

In the small chemical potential region before the phase transition as shown in panel~(a), the dynamical up-quark mass without the quark AMM ($\bar v_{u,d}=0$) increases as the magnetic field strengthens. This behavior is known as the magnetic catalysis for the chiral symmetry breaking.
Once we include the quark AMM ($\bar v_{u,d}=1\,{\rm GeV}^{-3}$), this enhancement is suppressed by the quark AMM. As the AMM parameter increases to $\bar v_{u,d}=2\,{\rm GeV}^{-3}$,
the dynamical up-quark mass tends to decrease with increasing the magnetic field.
Furthermore, when we consider the case of opposite signs for the AMM parameters ($\bar v_u=2\,{\rm GeV}^{-3},\; \bar v_d=-2\,{\rm GeV}^{-3}$), 
there is no significant change compared to the case of $\bar v_{u,d}=2\,{\rm GeV}^{-3}$.

Moving onto the large quark chemical potential regions after the phase transition, 
panel~(a) shows that
the quark chemical potential dependence of the dynamical up-quark mass is not significantly affected by the quark AMM contribution for all magnetic fields. Since the quark AMM is associated with the quark condensate, its contribution becomes negligible in the chiral symmetric phase.

Panel~(b) illustrates the same as panel~(a), but for the dynamical down-quark mass. 
The results are almost similar to those of the dynamical up-quark mass except in the case of $\bar v_u=2\, {\rm GeV}^{-3},\; \bar v_d=-2\,{\rm GeV}^{-3}$. In this case,
the dynamical down-quark mass increases with increasing the magnetic field, which is in contrast to the dynamical up-quark mass.

\begin{figure}[H] %htbp
\begin{center}
    \includegraphics[width=7.8cm]{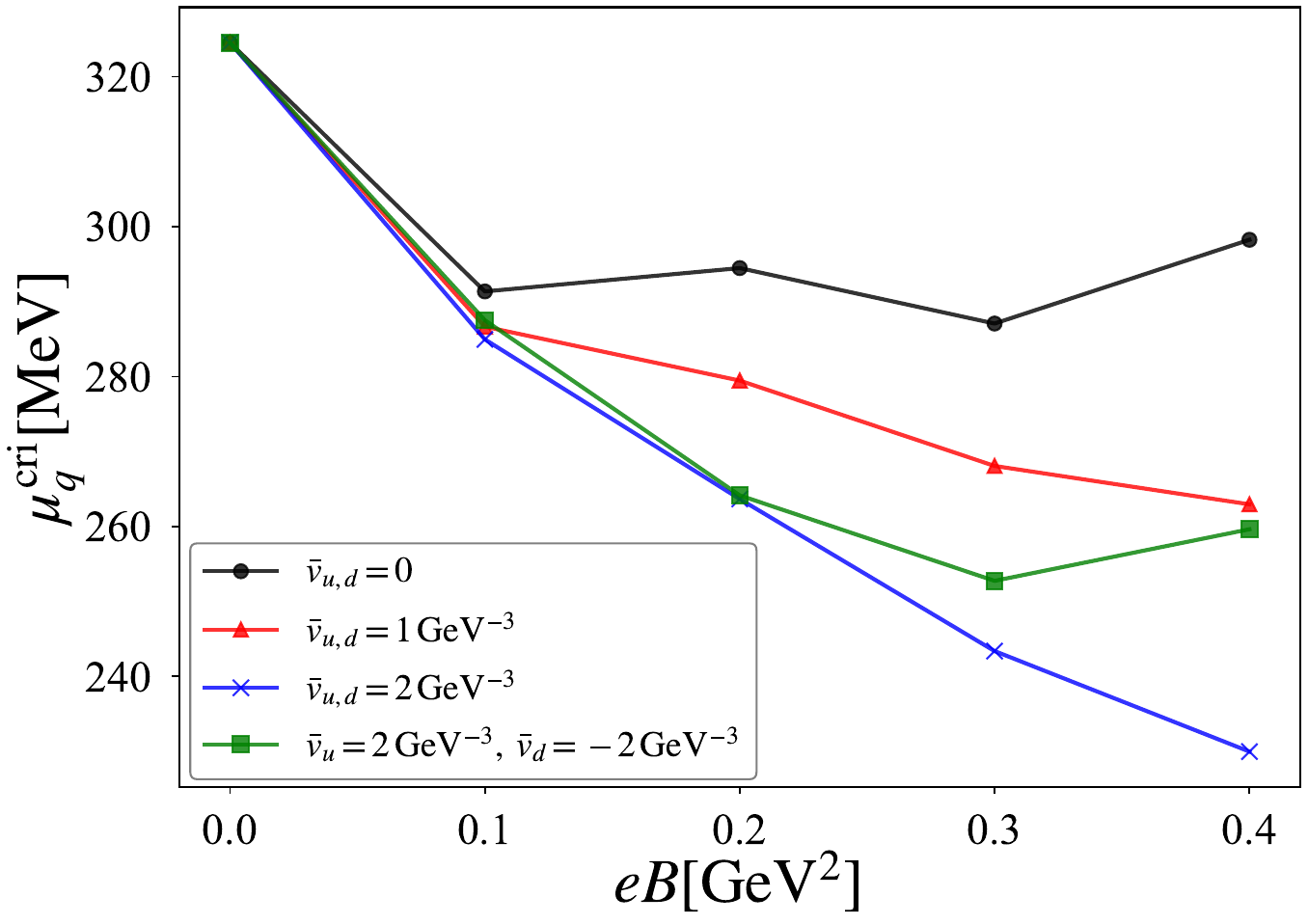}
    %\subfigure{(a)}
\end{center}
\caption{
Magnetic effect on the critical quark chemical potential influenced by the quark AMM
}
\label{critical_muq}
\end{figure}

We read the critical quark chemical potential $\mu_q^{\rm cri}$
corresponding to the phase transition point
from Fig.~\ref{mass_chemi} and 
plot its value in Fig.~\ref{critical_muq}.
In the absence of the magnetic field, 
the dynamical quark masses smoothly decrease with increasing quark chemical potential, resulting in
a chiral crossover. 
The pseudocritical chemical potential of this chiral crossover is estimated as $\mu_q^{(\rm pc)}=324.5\,{\rm MeV}$
\footnote{
Using another parameter set,
$\Lambda = 569.52{\rm MeV}$,
$m_u=m_d= $ 5.455{\rm MeV},
$G_S \Lambda^2= 2.333$, one can observe the first-order phase transition. However, recent analyses of neutron star (without the magnetic field) based on observational constraints from neutron star mass and radius
relations
favor the quark-hadron continuity, which implies the chiral crossover
(for details, see Ref.~\cite{Kojo:2020krb}).
To align with the quark-hadron continuity scenario, we adopt the parameter set of providing the crossover.
}.

When the external magnetic field is present,
the chiral phase transition changes to the first order.
Without the quark AMM contribution, the critical chemical potential decreases due to the presence of the external magnetic field. 
Including the quark AMM further decreases
the critical chemical potential. In particular, in the case of $\bar v_{u,d}=2\,{\rm GeV}^{-3}$ with the strong magnetic field $eB=0.4{\rm GeV^2}$,
$\mu_q^{\rm cri}$ is significantly decreased.
This indicates that
the quark AMM plays a role in promoting the chiral restoration in the neutral dense quark matter under the magnetic field.
Incidentally, such behavior has also been observed in the isospin symmetric system ($\mu_u=\mu_d$) in~\cite{He:2022inw}, where the two-flavor NJL model is employed with the quark AMM but without chemical equilibrium condition (Eq.~\ref{beta_equ}) or the electric charge neutrality condition (Eq.~\ref{neutrality}).

%%%%%%%%%%%%%%%%%%%%%%%%%%%%%%%%%%%%%%%%%%%%%%%%%%%
\subsection{Quark number density}

\begin{figure}[H] %htbp
\begin{tabular}{cc}
\begin{minipage}{0.5\hsize}
\begin{center}
    \includegraphics[width=8.8cm]{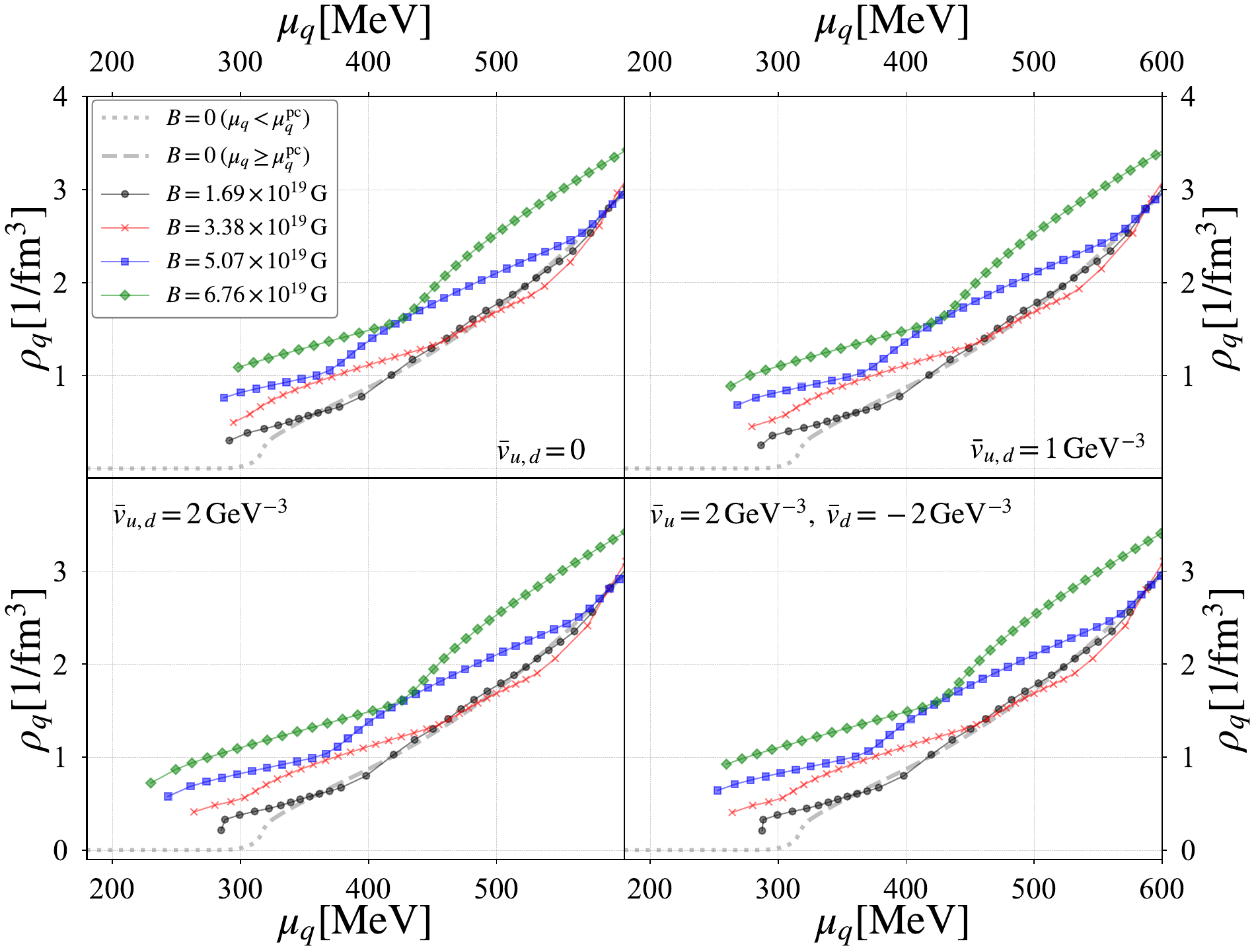}
    \subfigure{(a)}
\end{center}
\end{minipage}
\begin{minipage}{0.5\hsize}
\begin{center}
    \includegraphics[width=6.8cm]{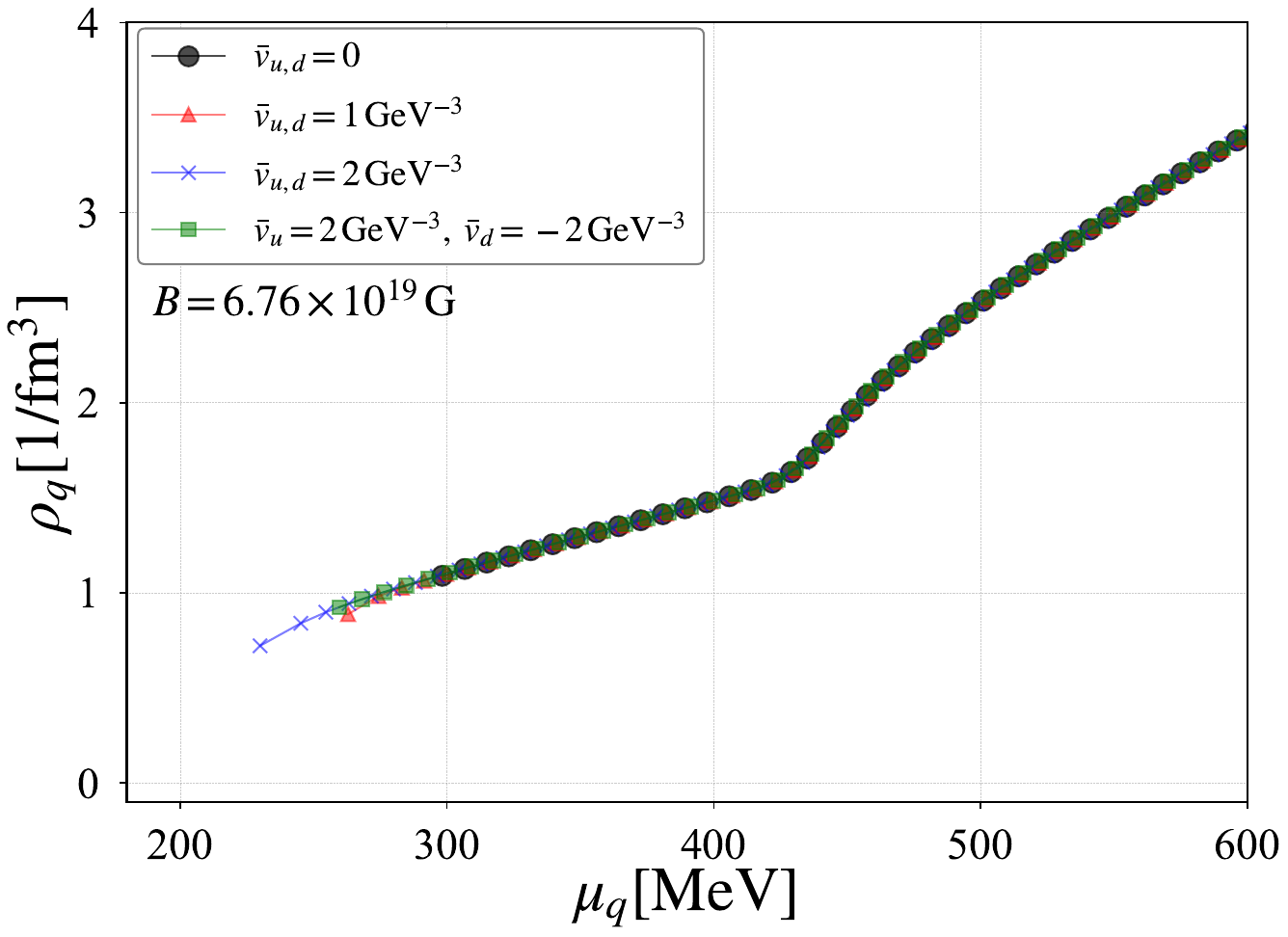}
    \subfigure{(b)}
\end{center}
\end{minipage}
\end{tabular}
\caption{
Quark number density as the function of the quark chemical potential. 
Panel~(a) shows that 
quark number density for each magnetic fields in all cases of the quark AMM contribution. Here, for $eB=0$,
the dotted plot corresponds to the quark number density before the pseudocritical quark chemical potential, while the dashed line represents the density after the pseudocritical quark chemical potential. 
Panel~(b) focuses on 
$B = 6.76\times 10^{19}\,{\rm G}$
%$eB=0.4\,{\rm  GeV}^2$
and alternatively presents the quark AMM contribution.
Note that the connection between
${\rm GeV}^2$ and ${\rm G}$ is interpreted by using the following correspondence:
$B = 1.69\times 10^{19}\,{\rm G}$ corresponds to
$eB=0.1\,{\rm GeV^2}$.
}
\label{density}
\end{figure}

In this subsection, we show the numerical results for the quark number density as a function of the quark chemical potential in Fig.~\ref{density}. In the absence of the magnetic field, the quark number density begins to smoothly increase with increasing the quark chemical potential, owing to the chiral crossover.  
In the presence of the magnetic field, the first-order chiral phase transition occurs as shown in Fig.~\ref{mass_chemi}, resulting in a jump in the quark number density at the critical quark chemical potential.
After the critical chemical potential, the quark number density suddenly gets a finite value.
Within the NJL model framework,
the quark matter phase is realized after the first-order chiral phase transition.

Looking at the case without the quark AMM in panel~(a),
the quark number density tends to be enhanced by the magnetic field. 
When the quark AMM is included,
the quark number density near the critical quark chemical potential take smaller values compared to the case in the absence of the magnetic field. 
Moving onto the high-density regions, the quark AMM hardly contributes to the quark number density.
To clearly observe this, we show panel~(b) which is different representation of the quark AMM contribution to the quark number density. Here, the magnetic field is fixed at %\textcolor{red}{
$B =6.76\times 10^{19}\,{\rm G}$
(corresponding to $eB=0.4\,{\rm GeV^2}$)
%}
and the quark AMM parameter is varied.  
This figure surely shows that 
the influence of
the quark AMM is negligible in the high-density regions. However it contributes mainly to low-density regions up to the critical quark chemical potential.
This is because the quark AMM interaction vanishes with the chiral restoration.
Similar behavior is observed for different magnetic fields.
Note that the role of the quark AMM is comparable in the isospin symmetric analysis, where chemical equilibrium condition (Eq.~\ref{beta_equ}) and the electric charge neutrality condition (Eq.~\ref{neutrality}) are not applied, as was discussed in~\cite{He:2022inw}.

Note that in the low-density region 
near the critical chemical potential, the quark number density is occupied by the lowest Landau level (LLL), as was investigated in~\cite{He:2022inw}. 
When the quark chemical potential increases, 
the excited states tagged with 
the higher landau levels, which have
the heavier effective quark masses,
begins to contribute to the effective potential as well as
the quark number density.

%%%%%%%%%%%%%%%%%%%%%%%%%%%%%%%%%%%%%%
\subsection{EoS without magnetization}

\begin{figure}[H] %htbp
\begin{center}
    \includegraphics[width=12.8cm]{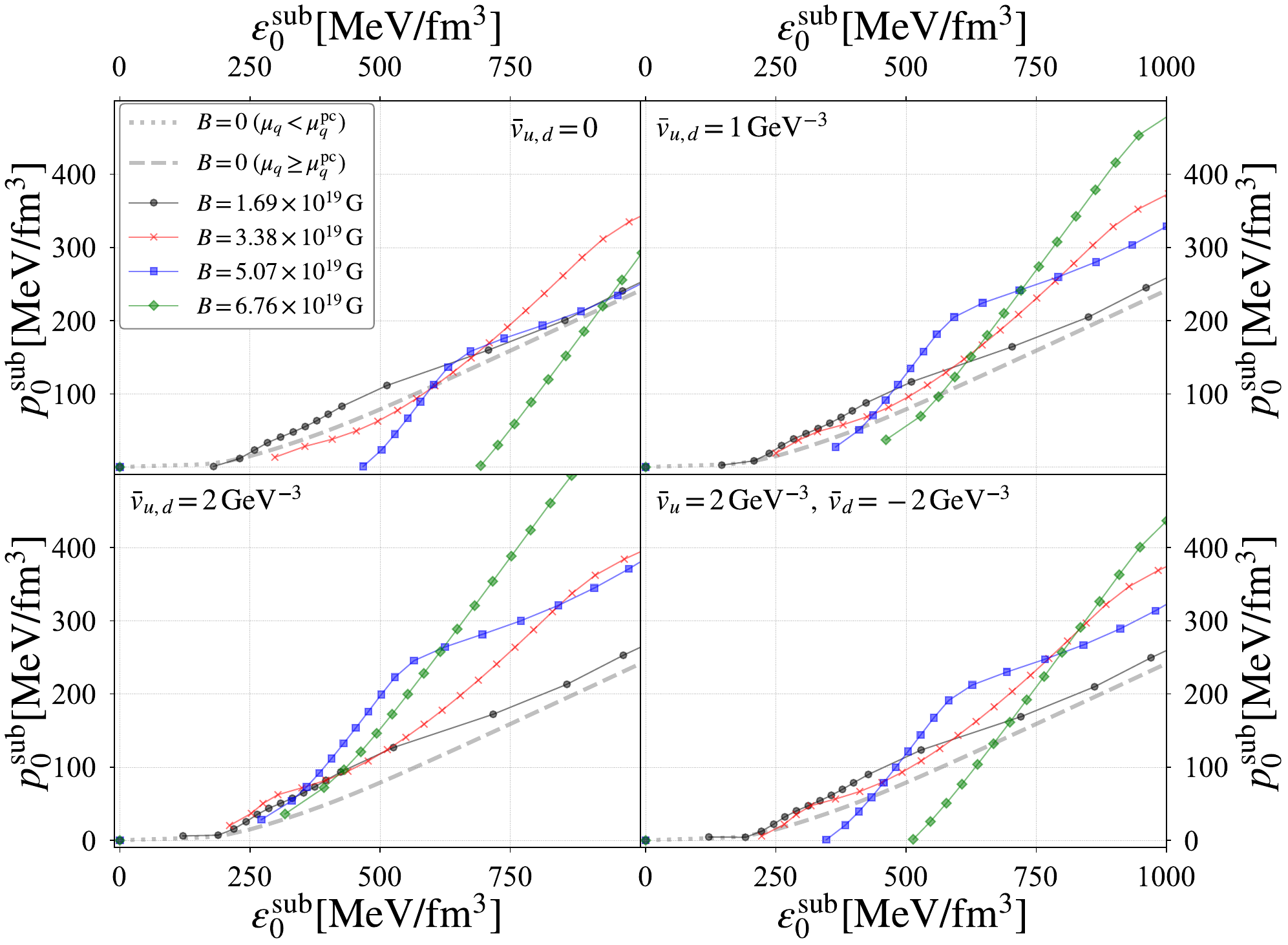}
    %\subfigure{(a)}
\end{center}
\caption{
Quark AMM contribution to pressure as a function of energy.  
Here, the magnetization is not taken into account.
}
\label{E_P_wo_magne}
\end{figure}

In this subsection, we discuss the effect of the quark AMM on the isotropic EoS described by the pressure and energy, where the magnetization is not included.
Figure~\ref{E_P_wo_magne} plots
the isotropic pressure as a function of the energy.
First, we focus on the case without the quark AMM.
For $B = 1.69\times 10^{19}\,{\rm G}$
(corresponding to $eB=0.1{\rm GeV}^2$),
the pressure is somewhat larger than
that in the absence of the magnetic field.
This behavior is explained as follows.
Looking at the critical chemical potential
where the pressure approaches zero,
the energy is dominated by the up- and down-quark number densities with chemical potentials, $\left.\epsilon_{0}^{\rm sub}\right|_{\mu_q=\mu_q^{\rm cri}}\sim \sum_{i=u,d} \mu_i^{\rm cri} \rho_i^{\rm cri}$.
Indeed, by including the magnetic field 
$B = 1.69\times 10^{19}\,{\rm G}$, 
%$eB=0.1\,{\rm GeV}^2$,
the critical quark chemical potential becomes smaller than the pseudocritical quark chemical potential in the absence of the magnetic field (see Fig.~\ref{critical_muq}), whereas the value of the quark number density almost remains steady at the critical quark chemical potential,
as can be inferred from Fig.~\ref{density}.
Therefore, the energy at the critical quark chemical potential is smaller than that in the absence of the magnetic field. As the energy increases from its value at the critical chemical potential, the pressure begins to increase as well. Consequently, the pressure for $B = 1.69\times 10^{19}\,{\rm G}$
%$eB=0.1{\rm GeV^2}$
is always higher than when $B=0$, resulting in stiffer 
EoS of neutral dense quark matter due to the presence of a magnetic field with 
$B = 1.69\times 10^{19}\,{\rm G}$.
%$eB=0.1\,{\rm GeV}^2$

As the magnetic field increases, the energy at the critical quark chemical potential shifts to a higher value. This is due to the increase in the quark number density at the critical chemical potential,
while the critical quark chemical potential itself does not change drastically, as depicted in Fig.~\ref{density}. 
As the energy starts to increase, the pressure begins to rise but remains smaller than that in the absence of the magnetic field, suggesting that the strong magnetic field softens the EoS of the neutral dense quark matter.
However, the energy further increases, the pressure eventually exceeds that in the absence of the magnetic field.

Next we consider the quark AMM contribution to the energy and pressure.
For 
$B = 1.69\times 10^{19}\,{\rm G}$,
%$eB=0.1\,{\rm GeV^2}$,
the energy and pressure are hardly affected by the quark AMM contribution. However, as the magnetic field increases, the AMM contribution become significant.
Indeed, for 
$B = 3.38,5.07,6.76\times 10^{19}\,{\rm G}$
%$eB=0.2,0.3,0.4\,{\rm GeV^2}$,
the plot of the energy and pressure shifts to the left in Fig.~\ref{E_P_wo_magne}: the neutral dense quark matter can have  pressure at lower energy under the magnetic fields, owing to the quark AMM.
This is because the critical quark chemical potential is decreased by the quark AMM and the quark number density at the critical quark chemical potential
is  correspondingly reduced.
This effect is particularly pronounced 
in the case of the large quark AMM $\bar v_{u,d}=2\,{\rm GeV}^{-3}$.

%%%%%%%%%%%%%%%%%%%%%%%%%%%%%%%%%%%%%%%%%%%%%%%%%%%
\subsection{Subtracted magnetization}

In this subsection, we discuss the magnetization in the neutral dense quark matter. Since the magnetization suffers from the ultraviolet divergence, we use the subtracted magnetization in Eq.~(\ref{sub_magne}).

\begin{figure}[H] %htbp
\begin{center}
    \includegraphics[width=12.8cm]{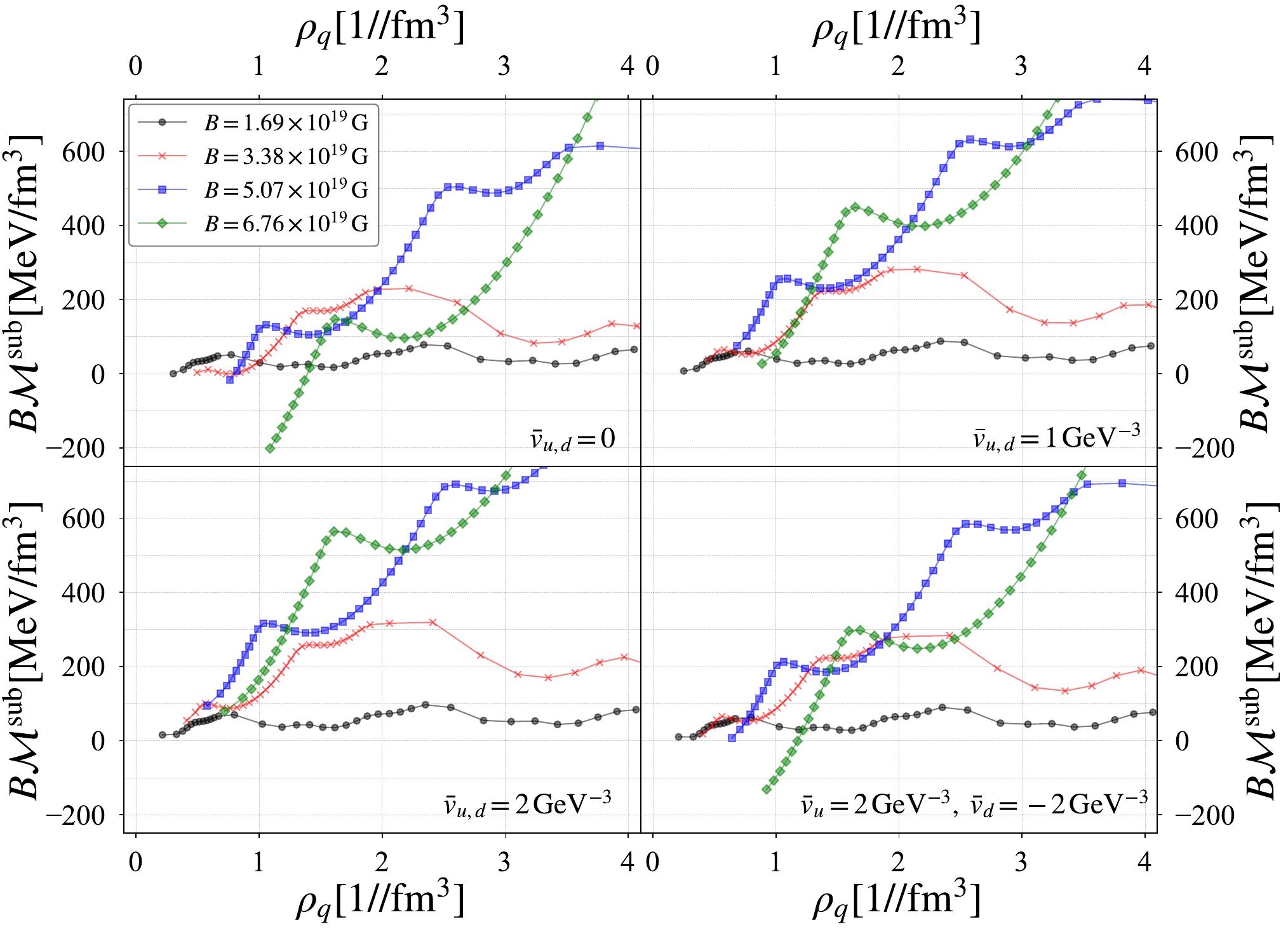}
    %\subfigure{(a)}
\end{center}
\caption{
Subtracted magnetization as a function of the quark number density.  
}
\label{magnetization}
\end{figure}

Figure~\ref{magnetization} plots 
the quark number density dependence of
the subtracted magnetization, 
where the magnetization is multiplied by the magnetic field.
In the absence of quark AMM, for
$B = 1.69\times 10^{19}\,{\rm G}$,
%$eB=0.1\,{\rm GeV}^2$,
the subtracted magnetization takes on a positive finite value and remains small for all the quark chemical potential.
As the magnetic field increases,
the strength of magnetization tends to be enhanced. In particular, 
for 
$B = 5.07\times 10^{19}\,{\rm G}$,
%$eB=0.3\,{\rm GeV}^2$, 
the magnetization significantly enlarges in the high density region.

In contrast to the regions with
the small magnetic field, for the strong magnetic field of 
$B = 6.76\times 10^{19}\,{\rm G}$, the magnetization takes on a negative value in the low density region near the critical chemical potential, indicating the diamagnetism.
The detail of the sign change of the magnetization around the critical quark chemical potential
is understood as follows. 
The subtracted magnetization can be rewritten as 
\begin{eqnarray}
B{\cal M}^{\rm sub}
&=&
V_{\rm S}^{\rm sub}
+
p_0^{\rm sub}
+
B {\cal M}^{\rm sub}_{\rm loop}
\end{eqnarray}
where 
\begin{eqnarray}
V_{\rm S}^{\rm sub}&=&
V^{\rm S}_{\rm eff} (\mu_u,\mu_d,\mu_e, \mu_\mu) 
-
V^{\rm S}_{\rm eff} (\mu_u=\mu_d=\mu_e= \mu_\mu=0),\nonumber\\
{\cal M}^{\rm sub}_{\rm loop}
&=&
-B\frac{\partial}{\partial B}\left(
\frac{V^{\rm loop}_{\rm eff} (\mu_u,\mu_d,\mu_e, \mu_\mu)    }{B}
-
\frac{V^{\rm loop}_{\rm eff} (\mu_u=\mu_d=\mu_e= \mu_\mu=0)    }{B}
\right).
\end{eqnarray}
At the critical quark chemical potential,
the subtracted pressure without the quark AMM 
is negligibly small,
$\left.
p_0^{\rm sub}
\right|_{\mu_q=\mu_q^{\rm cri}}\sim0$.
Hence, the subtracted magnetization can be expressed as
\begin{eqnarray}
\left.
B{\cal M}^{\rm sub}\right|_{\mu_q=\mu_q^{\rm cri}} 
&\simeq&
%\left. V^{\rm S}_{\rm sub} \right|_{\mu_q=\mu_q^{\rm cri}}  - \left. B^2\frac{\partial}{\partial B}\left( \frac{V^{\rm loop}_{\rm sub}}{B} \right)\right|_{\mu_q=\mu_q^{\rm cri}} \nonumber\\
%&=&
\left.
V_{\rm S}^{\rm sub}
\right|_{\mu_q=\mu_q^{\rm cri}}
+
\left.
B {\cal M}^{\rm sub}_{\rm loop}\right|_{\mu_q=\mu_q^{\rm cri}}
\end{eqnarray}
The scalar mean-field potential part $V^{\rm S}_{\rm eff}$ is described by the chiral condensate $\sigma_{u,d}$ as shown in Eq.~(\ref{potential_detail}), which is associated with the strength of the spontaneous chiral symmetry breaking. Therefore, the subtracted scalar potential  $\left. V_{\rm S}^{\rm sub}\right|_{\mu_q =\mu_q^{\rm cri}}$ indicates the difference in the strength of the chiral symmetry breaking between the chiral symmetry broken phase and the chiral symmetry restored phase at $\mu_q =\mu_q^{\rm cri}$. This subtracted potential takes on
a negative value for any magnetic field, i.e., $\left. V_{\rm S}^{\rm sub}\right|_{\mu_q =\mu_q^{\rm cri}}<0$.
On the other hand, ${\cal M}_{\rm sub}^{\rm loop}$ is originated from the quark  and lepton loop contribution in the effective potential, which is dominated by the quark loop. This loop contribution is positive, 
$\left. B {\cal M}^{\rm sub}_{\rm loop}\right|_{\mu_q=\mu_q^{\rm cri}}
>0$.
The sign of the subtracted magnetization at the critical chemical potential is thus determined by the competition between the scalar mean-field potential and the loop contribution of quarks and leptons. 

For small magnetic fields, the magnitude of  the subtracted scalar potential is comparable to that of the quark loop contribution,
%$\left| \left. V^{\rm S}_{\rm sub} \right|_{\mu_q =\mu_q^{\rm cri}}\right| \sim\left| \left. B {\cal M}_{\rm sub}^{\rm loop}\right|_{\mu_q=\mu_q^{\rm cri}}\right|$
resulting in 
the marginally small subtracted magnetization
$\left.B{\cal M}^{\rm sub}\right|_{\mu_q=\mu_q^{\rm cri}}\sim0$.
For the strong magnetic field of 
$B = 6.76\times 10^{19}\,{\rm G}$,
%$eB=0.4{\rm GeV}^2$,
the magnitude of the subtracted scalar potential competes the loop contribution, leading to the negative subtracted magnetization, $\left.
B{\cal M}^{\rm sub}\right|_{\mu_q=\mu_q^{\rm cri}} <0$.

%\begin{eqnarray}
%\left.
%V^{\rm S}_{\rm sub}
%\right|_{\mu_q =\mu_q^{\rm cri}}&<&0,\nonumber\\
%\left.
%B {\cal M}_{\rm sub}^{\rm loop}\right|_{\mu_q=\mu_q^{\rm cri}}
%&>&0,\nonumber\\
%\left|
%\left.
%V^{\rm S}_{\rm sub}
%\right|_{\mu_q =\mu_q^{\rm cri}}\right|
%&>&\left|
%\left.
%B {\cal M}_{\rm sub}^{\rm %loop}\right|_{\mu_q=\mu_q^{\rm cri}}\right|
%\end{eqnarray}

When the quark AMM ($\bar v_{u,d} =1,2\,{\rm GeV}^{-3}$) is included, 
the subtracted pressure takes a nonzero positive value at the critical quark chemical potential,
$\left.
p_0^{\rm sub}
\right|_{\mu_q=\mu_q^{\rm cri}}>0$, which enlarges the subtracted magnetization to a positive value. 
However,
for the opposite AMM parameter ($\bar v_u =2\,{\rm GeV}^{-3},\; \bar v_d=-2\,{\rm GeV}^{-3}$), 
the subtracted pressure is negligible. Similar to the case without the quark AMM,
the scalar mean-field potential competing with the loop contribution provides the negative magnetization at the critical quark chemical potential under the strong magnetic field.
Furthermore, 
the strength of the magnetization is enhanced by any quark AMMs for the entire range of the quark chemical potential.

%%%%%%%%%%%%%%%%%%%%%%%%%%%%%%%%%%%%%%%%%%%%%%%%%%%
\subsection{EoS with magnetization}

\begin{figure}[H] %htbp
\begin{center}
    \includegraphics[width=12.8cm]{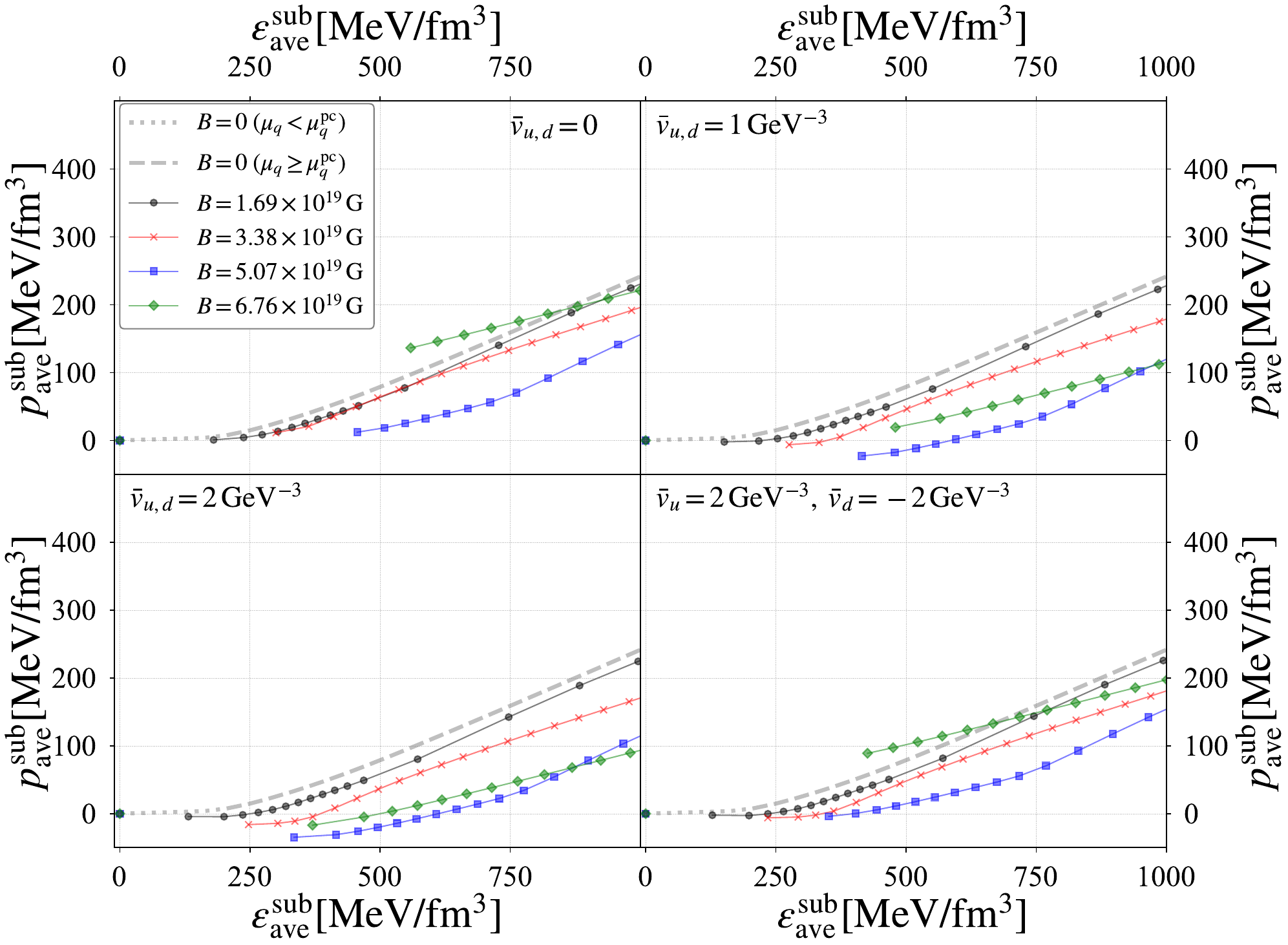}
    %\subfigure{(a)}
\end{center}
\caption{
Quark AMM contribution to the averaged pressure as a function of the averaged energy, incorporating the contribution of the magnetization.  
%Quark AMM contribution to the averaged energy and pressure as the function of the quark chemical potential.  
}
\label{E_P_w_magnetization}
\end{figure}

In this subsection, we take into account the magnetization in the energy and pressure.
Using the energy and pressure with and without the magnetization,
 we  will subsequently investigate the mass and radius relation, by solving the Tolman-Oppenheimer-Volkoff (TOV) equation~\cite{Tolman:1939jz,Oppenheimer:1939ne}:
\begin{eqnarray}
\frac{dM(r)}{dr} &=&4\pi r^2 \epsilon(r),\nonumber\\
\frac{dp(r)}{dr}&=&
-\frac{G\epsilon(r) M(r)}{r^2}\left[
1+ \frac{p(r)}{\epsilon(r)}
\right]
\left[
1+\frac{4\pi p(r) r^3}{M(r)}
\right]
\left[
1-\frac{2GM(r)}{r}
\right]^{-1},
\label{TOV}
\end{eqnarray}
where $M(r)$ is the total mass inside the spherical neutral dense quark matter with the radius $r$; the energy and pressure at $r$ denoted as $\epsilon(r)$ and $p(r)$, respectively;
$G$ is the gravitational constant.
The TOV equation is described based on spherically symmetric system.
However, including magnetization makes the pressure anisotropic, and it would be inappropriate to directly apply the pressure and energy including the magnetization to the TOV equation. 
Ideally, the TOV equation should be modified to take into account the anisotropies. 
Nevertheless, in this study, we define the following averaged pressure and energy to incorporate the magnetization, allowing us to solve the well-known TOV equations with these averaged quantities:
\begin{eqnarray}
 p^{\rm sub}_{\rm ave}
&=&\frac{
p_{\parallel}^{\rm sub} + 2p_{\perp}^{\rm sub}
}{3},\nonumber\\
\epsilon^{\rm sub}_{\rm ave} &=& - p^{\rm sub}_{\rm ave}  + 
\sum_{i=u,d,e,\mu} \mu_i \rho_i.
\end{eqnarray}
%In this subsection, we discuss the averaged energy and pressure.
Figure~\ref{E_P_w_magnetization} depicts the averaged energy and pressure incorporated with the magnetization.
Compared to the case without the magnetization in Fig.~\ref{E_P_wo_magne},
it is observed that the slope of pressure as a function of energy decreases
for all cases of the quark AMM,
due to the presence of the magnetization.
Even for 
$B = 1.69\times 10^{19}\,{\rm G}$,
%$eB=0.1\,{\rm GeV}^2$
the averaged pressure is lower than that in the absence of the magnetic field. 
Furthermore, for the strong magnetic field,
the sharp rise in the pressure at high energy, as observed in Fig.~\ref{E_P_wo_magne}, is also suppressed by the magnetization.
These results indicate that the magnetization softens the EoS for all cases of the quark AMM and for
all magnetic fields.

%%%%%%%%%%%%%%%%%%%%%%%%%%%%%%%%%%%%%%%%%%%%%%%%%%%%%%%%%%%%%%%%%%%%%%%%%%%%%%%%%%
\subsection{Mass and radius relation}

In this subsection, we evaluate the mass and radius relation of the neutral dense quark matter under the magnetic field 
through solving the TOV equation 
and examine the quark AMM contribution.

In order to evaluate the mass and radius in the neutral dense quark matter,
we adapt the energy and pressure in Fig.~\ref{E_P_wo_magne}  (the averaged one in Fig.~\ref{E_P_w_magnetization}) for the TOV equation. 
For numerically solving the TOV equation in Eq.~(\ref{TOV}), 
we apply the following initial conditions at the center of the spherical neutral dense quark matter, $r=0$:
we select a set of the energy and pressure from Fig.~\ref{E_P_wo_magne}  (the averaged one from Fig.~\ref{E_P_w_magnetization}) 
and assign these as the central values, $\epsilon_c = \epsilon(r=0)$ and $p_c = p(r=0)$. 
Additionally, the initial value of the mass is set to $M(r=0)=0$.
As the radius expands with solving the TOV equation, the 
$r$-dependent pressure decreases and reach its minimum value
corresponding to that at the critical chemical potential, 
$p(r=R)=\left. p\right|_{\mu_q=\mu_q^{\rm cri}}$. This radius $r=R$ is identified as the total radius of the neutral dense quark matter. In addition, the mass $M(r=R)$ is regarded as the total mass of the neutral dense quark matter.
To obtain a comprehensive set of results for the total mass and radius results,
we select different sets of the center values of the energy and pressure corresponding to varying quark chemical potentials, and repeat the above process.

\begin{figure}[H] %htbp
\begin{tabular}{cc}
\begin{minipage}{0.5\hsize}
\begin{center}
    \includegraphics[width=8.8cm]{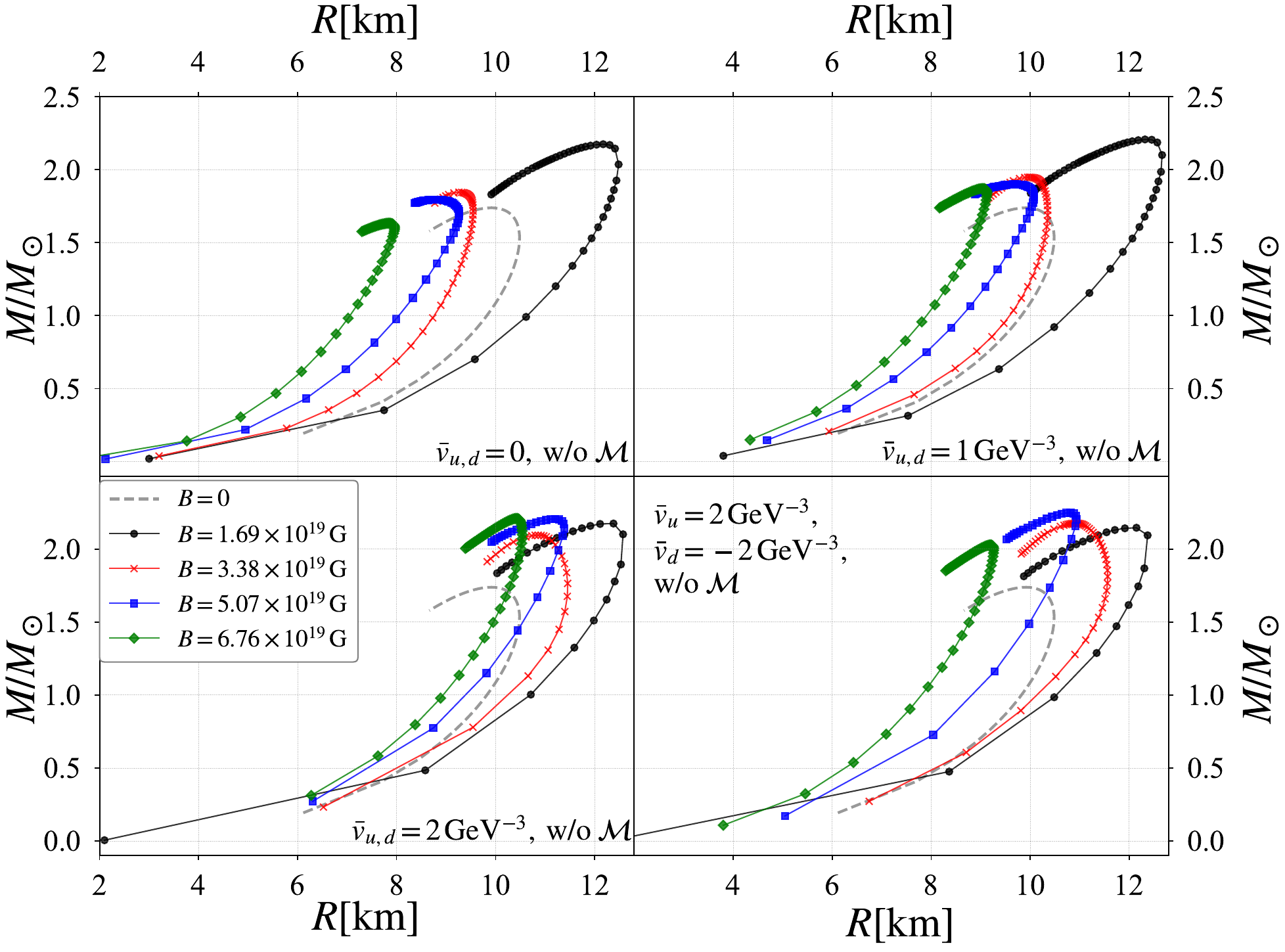}
    \subfigure{(a)}
\end{center}
\end{minipage}
\begin{minipage}{0.5\hsize}
\begin{center}
    \includegraphics[width=8.8cm]{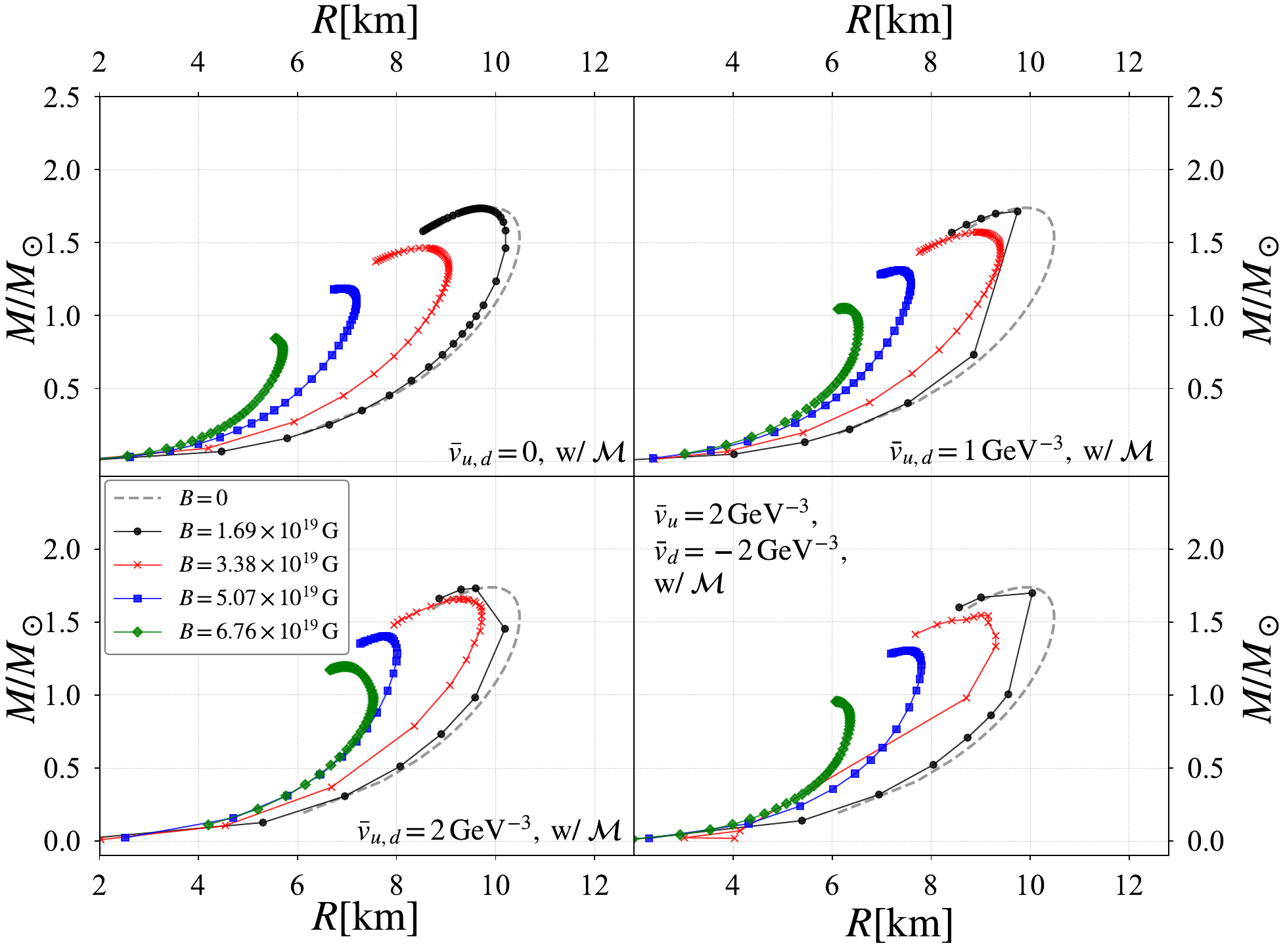}
    \subfigure{(b)}
\end{center}
\end{minipage}
\end{tabular}
\caption{
Quark AMM contribution to
mass-radius relation of the neutral dense quark matter:
Panel~(a) without the magnetization and Panel~(b)  with the magnetization.
$M_\odot$ denotes the solar mass.
}
\label{MR_relation}
\end{figure}

Figure~\ref{MR_relation} shows the mass-radius relation for neutral dense quark matter. Panel~(a) presents the case without magnetization, while panel~(b) includes the magnetization through the averaged pressure and energy.

First, we focus on panel~(a). Without  the quark AMM, the magnetic field of 
$B = 1.69\times 10^{19}\,{\rm G}$
%$eB=0.1\,{\rm GeV}^2$
enhances the mass and radius compared to the absence of the magnetic field.
This is because the magnetic field of 
$B = 1.69\times 10^{19}\,{\rm G}$
%$eB=0.1\,{\rm GeV}^2$ 
stiffens the EoS, as shown in the energy-pressure relation in Fig.~\ref{E_P_wo_magne}.
However, as the magnetic field increases without the quark AMM, the EoS tends to soften, especially in regions of not very high energy.
Therefore, the radius shrinks and the mass decreases with the increase in the magnetic field.

When including the AMM, for 
$B = 1.69\times 10^{19}\,{\rm G}$
%$eB=0.1\,{\rm GeV}^2$
in all cases of the quark AMM, the mass-radius relation is similar to the case without the quark AMM and remains larger than in the absence of the magnetic field. As the magnetic field increases, 
the quark AMM suppresses this softening caused by stronger magnetic fields, as shown in Fig.~\ref{E_P_wo_magne}. This effect of the quark AMM is reflected in the mass-radius relation.
In particular, for the case of $\bar v_{u,d}=2\,{\rm GeV}^{-3}$ and $\bar v_u=2\,{\rm GeV}^{-3},\; \bar v_d=-2\,{\rm GeV}^{-3}$, the radius shrinks while
the the maximum mass remains nearly the same as with 
$B = 1.69\times 10^{19}\,{\rm G}$.
%$eB=0.1\,{\rm GeV}^2$

However, when the magnetization is included, the EoS becomes softer for all cases of the quark AMM and for all magnetic fields, as depicted in Fig.~\ref{E_P_w_magnetization}. 
Indeed, panel~(b) of Fig.~\ref{MR_relation} shows that
as the magnetic field increases, the neutral dense quark matter becomes smaller for all cases of the quark AMM.
Furthermore, regardless of the sign of the magnetization,
the stronger magnetic field makes the neutral dense quark matter more compact and lighter.

%%%%%%%%%%%%%%%%%%%%%%%%%%%%%%%%%%%%%%%%%%%%%%%%%%%%%%%%%%%%%%%%%% 
\section{summary and discussion}

We have investigated the effect of the quark AMM on the EoS of the neutral dense quark matter under the magnetic field.
To anatomize the EoS under the magnetic field, we consider the cases with and without the magnetization.

In the case without the magnetization, the pressure is described as isotropic.
In the absence of the quark AMM,
the magnetic field of 
$B = 1.69\times 10^{19}\,{\rm G}$ (corresponding to $eB=0.1\,{\rm GeV^2}$)
%$eB=0.1\,{\rm GeV}^2$ 
stiffens the isotropic EoS. 
This stiffness can be understood in terms of the magnetic effect on the quark number density at the critical chemical potential. By applying the magnetic field of 
$B = 1.69\times 10^{19}\,{\rm G}$,
%$eB=0.1\,{\rm GeV}^2$
the critical quark chemical potential is smaller than in the absence of the magnetic field, while the quark number density at $\mu_q=\mu_q^{\rm cri}$ remains almost unchanged. 
Hence, in the plot of the energy versus pressure, the pressure begins to rise at a lower energy than $B=0$ scenario. As the energy increases, the pressure continues to grow and stays larger than $B=0$ scenario, resulting in the stiff EoS.
%Thus, the pressure for $eB=0.1\,{\rm GeV}^2$ is larger than when $eB=0$, resulting in the stiff EoS.
%As the magnetic field increases to
Moving to the stronger magnetic fields of 
$B = 3.38,\,
5.07,\,{\rm and}\, 6.76
\times 10^{19}\,{\rm G}$,
%$eB=0.2,0.3\,{\rm and}\, 0.4\,{\rm GeV}^2$, 
the quark number density at the critical chemical potential increases while the critical chemical potential remains almost the same as in the case of 
$B = 1.69\times 10^{19}\,{\rm G}$.
%$eB=0.1\,{\rm GeV}^2$.  
Then, the pressure begins to rise at a higher energy compared to $B=0$
and stays smaller than $B=0$ scenario
for regions of not very high energy. Consequently,
the EoS becomes softer with the increase in the magnetic field. Namely, the role of the stronger magnetic fields contrast with that of $B = 1.69\times 10^{19}\,{\rm G}$.
%$eB=0.1\,{\rm GeV}^2$.
In addition, this magnetic effect on the stiffness of the EoS is surely reflected in the mass-radius relation of the neutral dense quark matter. It is found that the mass-radius relation is increased by 
$B = 1.69\times 10^{19}\,{\rm G}$,
%$eB=0.1\,{\rm GeV}^2$,
but as the magnetic field further increases, the mass turns to decrease and the radius shrinks.

When the quark AMM is included, the critical chemical potential decreases, and the quark number density takes smaller values at the critical chemical potential.
However, for
$B = 1.69\times 10^{19}\,{\rm G}$,
%$eB=0.1\,{\rm GeV}^2$, 
the contribution of the quark AMM is not significant in the isotropic EoS and the mass-radius relation.
In contrast, the AMM contribution becomes more significant for the stronger magnetic fields. As a result, the quark AMM suppresses the softening effect on the EoS caused by the magnetic field of 
$B = 3.38,\,
5.07,\,{\rm and}\, 6.76
\times 10^{19}\,{\rm G}$,
%$eB=0.2,\,0.3,\,{\rm and}\,0.4\,{\rm GeV}^2$
leading to the increases in the mass and radius compared to the case of the respective magnetic field without the quark AMM. 

%Although the quark AMM contributes for all magnetic field regions, it is not significant for $eB=0.1\,{\rm GeV}^2$ in the isotropic EoS and the mass-radius relation.

Finally, we have taken the magnetization into account in the EoS of the neutral dense quark matter. 
To incorporate the magnetization into the pressure, we have defined the averaged pressure and energy for solving the spherical TOV equation. 
It is found that including the magnetization softens the EoS across all cases of quark AMM and magnetic fields, leading to decreases in both mass and radius.  The magnetic effect on the stiffness of the EoS discussed above is overshadowed by the contribution of the magnetization,
and the significance of the quark AMM also becomes invisible in the mass-radius relation.
\\

In closing,
we give comments of our findings and implications.

\begin{itemize}
\item
In our study, we take the flavor symmetry breaking into account
in the dynamical quark masses. 
In the chiral broken phase, the dynamical quark masses are enhanced by the magnetic field in the absence of the quark AMM. When including the quark AMM, the enhancement of the dynamical up-quark mass is suppressed. In particular, for the large AMM parameter ($\bar v_{u,d}=2\,{\rm GeV}^{-3}$), the dynamical up-quark mass decreases as the magnetic field increases. A similar behavior is observed in the dynamical down-quark mass except in the case of the opposite AMM parameter  ($\bar v_{u}=2\,{\rm GeV}^{-3}, \,\bar v_{d}=-2\,{\rm GeV}^{-3}$). In this case, the dynamical down-quark mass is enhanced by the magnetic field, which contrasts with the behavior of the dynamical up-quark mass. 
Looking at the quark matter phase, near the critical quark chemical potential, there is some difference between the dynamical up- and down-quark mass, but at the large chemical potential, both dynamical up- and down-quark masses approach zero.  
Although the dynamical quark masses surely exhibit the flavor symmetry breaking under the magnetic field,
the significance of this breaking does not manifest in the EoS and the mass-radius relation.

\item 
In the case without the quark AMM,
the negative magnetization appears in the small quark number density regions near the critical chemical potential. Indeed, the sign of the magnetization without the quark AMM is determined by the competition between the scalar potential contribution associated with the strength of the chiral symmetry breaking in the vacuum and the quark and lepton loop contribution.
For 
$B =6.76
\times 10^{19}\,{\rm G}$,
%$eB=0.4\,{\rm GeV}^2$, 
the scalar potential contribution competes with the quark and lepton loop contribution, resulting in the negative magnetization.
When including the quark AMM, 
the additional pressure contribution is driven and makes the magnetization positive in the case of $\bar v_{u,d}=1\,{\rm and}\,2\,{\rm GeV}^{-3}$. However, in the case of the opposite AMM parameter $\bar v_{u}=2\,{\rm GeV}^{-3}, \,\bar v_{d}=-2\,{\rm GeV}^{-3}$, 
the additional pressure contribution is not driven near the critical chemical potential and then
the magnetization takes negative values.
Although the sign of the magnetization is a characteristic magnetic property of the neutral dense quark matter, 
the magnetization makes the EoS soft regardless of whether it takes the negative value.
For this reason, 
it is hard to discern  
the magnetic property from the mass-radius relation, as shown in panel~(b) of Fig.~\ref{MR_relation}.

\item
The quark AMM is actually expected to be one of the candidates for explaining the generation mechanism of a strong magnetic field in magnetars, which is evaluated in the spontaneous magnetization. However, our results on the magnetization in Fig.~\ref{magnetization} have shown that the quark AMM contribution is negligible even
for $B = 1.69\times 10^{19}\,{\rm G}$.
Our results may imply that the quark AMM contribution is marginal to
the spontaneous magnetization. 
%This presents a challenge of finding another mechanism of spontaneously generating a strong magnetic field within effective model approaches.

\item 
Following Ref.~\cite{Ferrer:2010wz}, we have introduced the longitudinal and transverse pressures. However, as pointed out by a previous study~\cite{Potekhin:2011eb},
the magnetization induces a bound current, which provides the additional contribution to the transverse pressure: $p_\perp + {\cal M}B$.
Thus, even in the presence of the external magnetic field, 
the contribution of the magnetization is canceled out in 
the EoS, which remains the isotropic pressure: $p_{\parallel} = p_\perp = p_0$. 
However, such an additional contribution is not provided in the current analysis within this mean-field approximation.
If the additional term were indeed included in our analysis, 
the results obtained in the case without the magnetization would be more accurate results than those with the magnetization.
To clarify this, it is worth exploring how the additional contribution of the magnetization is accounted for the NJL model analysis.

\item
In this study, to understand the role of the quark AMM in the neutral dense quark matter and its correlation with the chiral symmetry under the magnetic field, we consider an idealized situation: the simplified NJL model with the two quark flavors under constant magnetic fields. 
%\textcolor{red}{
Our constant magnetic field strength is somewhat larger than the expected values in the realistic situation where maximum expected estimation is of $O(10^{18}\, G)$.
Indeed, the influence of the quark AMM becomes prominent starting from
$B= 5.07 \times 10^{19}\,G\sim O(10^{20}\, G)$, which far exceeds the current estimates for magnetars.
Looking at the lowest magnetic field displayed in our figures, $B= 1.69 \times 10^{19}\, G\sim O(10^{19}\, G)$, the quark AMM contribution does not significantly affect the EoS, magnetization, or mass-radius relation. This implies that
when considering the currently expected magnetic field strengths in magnetar, the quark AMM contribution is negligible.
However this evaluation is based on the simplified model analysis. 
To precisely evaluate the AMM contribution, 
it is necessary to extend the analysis to a more realistic model describing the magnetar.
%}
%To achieve a more realistic model describing the magnetar, it is necessary to extend the analysis to the three quark flavors and combine a nucleon model with the NJL model. 

\item
%\textcolor{red}{
To create a more realistic situation, the strength of the magnetic field in the magnetar needs to depend on the quark (baryon) number density. This dependence should  also be taken into account in the model analysis.
The conventional density dependent magnetic field was suggested by Ref.~\cite{Bandyopadhyay:1997kh}. 
Additionally, the modern application of density-dependent magnetic fields was discussed in Refs.~\cite{Dexheimer:2016yqu,Avancini:2017gck}, which is described by the chemical potential. 
We have applied the modern expression of the density-dependent magnetic field to our model. However, we have encountered issues such as negative dynamical quark mass and negative electron density, which are nonphysical results.
Addressing the details of these issue is beyond the scope of our study, but
we will investigate them further in the future.
%}
%We leave these extensions for future studies.

\item
In our study, we utilized the effective form of the quark AMM as defined in Eq.~(\ref{AMM_form}). However, this expression is a phenomenological approximation made to align with lattice QCD observations~\cite{Lin:2022ied,Kawaguchi:2022dbq}. To provide a more concrete discussion,
it is essential to 
determine a more precise  expression of the quark AMM, which is linked to the nonperturbative aspects of QCD, specifically the spontaneous chiral symmetry breaking. A more rigorous approach may involve employing Dyson-Schwinger equations (DSE), which allow for a nonperturbative evaluation of the quark AMM.

\end{itemize}

\section*{ACKNOWLEDGMENT}
I.S. is supported by the Ministry of Science and Technology(MOST) of China under Grant No. QN2023205001L and RFIS-NSFC under Grant No. 12350410364. M.H. is supported in part by the National Natural Science Foundation of China (NSFC) Grant  Nos: 12235016, 12221005, and supported by the Strategic Priority Research Program of Chinese Academy of Sciences under Grant No XDB34030000, and the Fundamental Research Funds for the Central Universities.

%%%%%%%%%%%%%%%%%%%%%%%%%%%%%%%%%%%%%%%%%%%%%%%%%%%%%%%
%\bibliography{reference}

\end{document}